\documentclass[%
superscriptaddress,
amsmath,amssymb,
 aps, prd,
 twocolumn,
 tightenlines,
 nofootinbib
]{revtex4-1}

\usepackage{graphicx,url,amssymb,amsmath,rotating,color,units,wasysym,epsfig,hyperref,epstopdf}

\begin{document}

\title{Effects of calibration uncertainties on the detection and parameter estimation of isotropic gravitational-wave backgrounds}

\author{Junaid Yousuf}
\affiliation{Department of Physics, University of Kashmir, Srinagar 190006, India}
\author{Shivaraj Kandhasamy}%
\affiliation{Inter University Center for Astronomy and Astrophysics, Pune 411007, India}
\author{Manzoor A Malik}
\affiliation{Department of Physics, University of Kashmir, Srinagar 190006, India}

\date{\today}

\begin{abstract}
Gravitational-wave backgrounds are expected to arise from the superposition of gravitational wave signals from a large number of unresolved sources and also from the stochastic processes that occurred in the Early universe. So far, we have not detected any gravitational wave background, but with the improvements in the detectors' sensitivities, such detection is expected in the near future. The detection and inferences we draw from the search for a gravitational-wave background will depend on the source model, the type of search pipeline used, and the data generation in the gravitational-wave detectors. In this work, we focus on the effect of the data generation process, specifically the calibration of the detectors' digital output into strain data used by the search pipelines. Using the calibration model of the current LIGO detectors as an example, we show that for power-law source models and calibration uncertainties $\lesssim 10 \%$, the detection of isotropic gravitational wave background is not significantly affected. We also show that the source parameter estimation and upper limits calculations get biased. For calibration uncertainties of $\lesssim 5 \%$, the biases are not significant ($\lesssim 2 \%$), but for larger calibration uncertainties, they might become significant, especially when trying to differentiate between different models of isotropic gravitational-wave backgrounds. 

\end{abstract}

\maketitle

\section{Introduction}
Since the first detection in September 2015 \cite{GW150914}, the LIGO \cite{aLIGO_2015}, and the Virgo \cite{Virgo_2015} gravitational wave (GW) detectors have detected nearly one-hundred compact binary merger signals \cite{GWTC-1, GWTC-2, GWTC-3}. They correspond to individual merger signals with a high signal-to-noise ratio (SNR). In addition to those high SNR signals, assuming the merger events are outliers in a much larger population of compact mergers, we also expect many low SNR signals that are hard to detect individually. The superposition of such a large number of low SNR signals would give rise to a gravitational-wave background (GWB) that could be detected with the current or next generation of GW detectors \cite{GW150914_stoch, gw170817_stoch, ratesandpop_GWTC3, tania_symmetry22}. 

Apart from the compact binary mergers signals, superposition of other astrophysical GW signals such as from core-collapse supernovae \cite{PhysRevD.72.084001, 2004MNRAS.351.1237H}, magnetars \cite{2012PhRvD..86j4007R, 2011ApJ...729...59Z} could also give rise to GWB. In addition to these astrophysical sources, various events that took place in the early universe such as inflation and phase transitions could also give rise to GWB \cite{SGWBCaprini_2018}. The detection of GWB from astrophysical sources can help us better understand the population and the evolution of stars in the universe \cite{SGWBRegimbau_2011, 2020arXiv200109663D, RomanoCornish_2017} while the detection of GWB from cosmological sources can provide information about the processes in the very early universe which are otherwise difficult to obtain \cite{2000PhR...331..283M}.

The LIGO-Virgo-KAGRA (LVK) collaboration, in their recent analyses using data from the observing run O3, did not find any evidence of GWBs and hence placed upper limits on the amplitudes of possible isotropic \cite{SGWBLVK_O3} and anisotropic GWBs \cite{SGWBLVK_dir_O3}. With the proposed improvements to the current GW detectors \cite{prospects_LRR}, it might be possible to detect the GWB from compact binary mergers \cite{tania_symmetry22}. Also, the proposed next-generation GW detectors \cite{cosmic_explorer, ET} are expected to observe the GWB from compact binary mergers with high SNRs \cite{sgwb_predict_3rd_1, sgwb_predict_3rd_2}. The data generation and various aspects of the search are expected to affect the GWB search results, and hence it is important to understand them. In this paper, we focus on the effects of the data generation, specifically that of the calibration, on the analysis results. Calibration is the process of converting the raw digital outputs of the detectors into strain data that are further used in the GW analyses. Any uncertainties in that process could translate into biases and larger uncertainties in the final results, affecting our interpretations. 

Typically, cross-correlation-based searches correlating data from multiple detectors are used to detect GWBs \cite{AllenRomano_1999}. In previous such searches using LIGO-Virgo data \cite{SGWBLVK_O1, SGWBLVK_O2, SGWBLVK_O3}, upper limits were calculated after marginalizing over calibration uncertainties as outlined in \cite{Whelan_2014}. However, that method does not capture any biases introduced by uncertainties and systematic errors in the calibration model. In this work, we try to address that issue. In the past, this has been studied primarily in the context of the search for GW signals from individual compact binaries \cite{allen_cal, salvo12, farr_2014, Hall_2019}. Recently, such questions have also been addressed for the detection and parameter estimation of individual compact binary merger signals \cite{Ethan20, Salvo21, Reed22}. We use a similar simulation-based method \cite{Ethan20, Salvo21} to address the effects of calibration uncertainties on the searches for GWB. In addition, we also show that one could try to estimate the GWB and calibration model parameters simultaneously and get a reasonable signal recovery.

The remainder of this paper is organized as follows. In Sec.~\ref{sgwb_model}, we briefly introduce the model and search for GWB using data from GW detectors. In Sec.~\ref{sec:calibration_model}, we discuss the calibration model used to convert the raw digital output into strain data used in GW searches. In Sec.~\ref{sec:analysis_method}, we describe the method used to quantify the effects of calibration uncertainties on the isotropic GWB searches. In Sec.~\ref{sec:results}, we show the results of our analyses, and in Sec.~\ref{sec:conclusions} conclude with the main results and future outlook.

\section{Modeling and search for isotropic gravitational-wave backgrounds} \label{sgwb_model}
An isotropic GWB is usually characterized in terms of fractional energy density in gravitational waves $\Omega_{gw}(f)$ \cite{AllenRomano_1999}, given by, 
\begin{equation}
   \Omega_{gw}(f)=\frac{f}{\rho_c}\frac{d\rho_{gw}}{df} \ ,
\end{equation}
where $f$ is the frequency, $d\rho_{gw}$ is the energy in gravitational waves in the frequency interval from $f$ to $f+df$, $\rho_c$ is the critical energy density needed to close the universe. The value of $\rho_c$ is given by
\begin{equation}
    \rho_c=\frac{3c^2H_0^2}{8\pi G} \ ,
\end{equation}
where $c$ is the speed of light, $G$ is the gravitational constant and $H$ is the Hubble constant. In this work, we use the value of Hubble constant measured by the Plank satellite, $H_0 = 67.9 \ {\rm km \ s^{-1} \ Mpc^{-1}}$ \citep{Planck_2015}. However, the conclusions drawn are independent of the actual value of $H_0$.

Typically $\Omega_{gw} (f)$ is expressed in the form of a power law,
\begin{equation}
    \label{eq:power_law}
  \Omega_{gw}(f) = \Omega_\alpha \left(\frac{f}{f_{\rm ref}}\right)^\alpha \ ,
\end{equation}
where $f_{\rm ref}$ is a reference frequency. For results reported in this paper, we use a reference frequency of $f_{\rm ref} = 25 \ {\rm Hz}$ as used in the LVK analyses \citep{SGWBLVK_O1, SGWBLVK_O2, SGWBLVK_O3}. The value of the power-law index $\alpha$ depends on the source of GWB we are interested in. For 
cosmological GWB from inflationary scenarios, we typically expect $\alpha = 0$ \cite{SGWBCaprini_2018} while for astrophysical GWB from the superposition of many compact binary mergers signals $\alpha = 2/3$ \cite{SGWBRegimbau_2011}. Similar to LVK analyses \citep{SGWBLVK_O1, SGWBLVK_O2, SGWBLVK_O3}, in addition to $\alpha = 0$ and $\alpha = 2/3$, we also look at $\alpha = 3$ representing astrophysical GWB models such as from supernovae \cite{alpha3}.

Instead of searching for $\Omega_{gw}(f)$, traditionally, isotropic GWB searches try to estimate $\Omega_\alpha$ for different values of power-law index $\alpha$. The optimal estimator of $\Omega_\alpha$, for an isotropic GWB, at a time $t$ and at a frequency bin $f$ is given by \cite{SGWBH1H2, RomanoCornish_2017},
\begin{equation}\label{eq:Omega}
    \hat{\Omega}_\alpha (t; f) = \frac{2}{T} \frac{\Re[d^*_I(t; f) d_J(t; f)]}{\gamma_{IJ}(f) S_\alpha(f)} \ ,
\end{equation}
where $d_1(t; f)$ and $d_2(t; f)$ are short-time Fourier transforms of the strain data from the two detectors $(I,J)$ evaluated at time $t$, $T$ is the duration of the data segments used for Fourier transforms and $\gamma_{IJ}(f)$ is the normalized overlap reduction function for the given two detectors $(I,J)$. The function $S_\alpha(f)$ is proportional to the assumed spectral shape $\alpha$ and is given by \citep{SGWBH1H2, RomanoCornish_2017},
\begin{equation}
    S_\alpha(f) = \frac{3 H^2}{10 \pi^2} \frac{1}{f^3} \left(\frac{f}{f_{\rm ref}}\right)^\alpha
\end{equation}
In the weak-signal limit, the variance of $\hat{\Omega}_\alpha$ is given by  \cite{SGWBH1H2, RomanoCornish_2017},
\begin{equation}\label{eq:Variance}
    \sigma^2_{\hat{\Omega}_\alpha} (t; f) = \frac{1}{2 T \Delta f} \frac{P_I(f) P_J(f)}{\gamma^2_{IJ}(f) S^2_\alpha(f)}
\end{equation}
where $P_I(f)$, $P_J(f)$ are the one-sided power spectral densities of the strain data from the two detectors $(I,J)$, and $\Delta f$ is the frequency resolution. For data spanning many segments and a large frequency band, the final optimal estimators are obtained by a weighted sum,
\begin{equation}\label{eq:combining_segments}
    \hat{\Omega}_\alpha = \frac{\sum_{t, f} \sigma_{\hat{\Omega}\alpha}^{-2}(t; f) \hat{\Omega}_\alpha(t; f)}{\sum_{t, f} \sigma_{\hat{\Omega}_\alpha}^{-2}(t; f)} , \ \ \
    \sigma_{\hat{\Omega}_\alpha}^{-2} = \sum_{t, f} \sigma_{\hat{\Omega}_\alpha}^{-2}(t; f) ,
\end{equation}
where $t$ runs over available time segments and $f$ runs over discrete frequency bins in the desired frequency band.

\section{Calibration Model}
\label{sec:calibration_model}
The raw outputs of gravitational wave detectors are digitized electrical signals from the photodetectors at the output port. The process of converting these electrical signals into strain data is called $\it calibration$. The LIGO, Virgo, and KAGRA  detectors have similar fundamentals in optical layout and control system topology \cite{aLIGO_2015, Virgo_2015, KAGRA_2021}. While their methods to describe and characterize that system are different (sometimes only in subtle ways that reflect their detailed differences), any of those methods could be used to describe current GW detectors. Thus, here, we follow and choose the methods of the LIGO detectors \cite{calibTechniqueLIGO_2015, Sun_2020}. For details of different calibration techniques used in the current generation of gravitational wave detectors, see \cite{calibTechniqueLIGO_2015, Viets_2018, Acernese_2022, calibKAGRA}. As shown in \cite{Sun_2020}, after detailed modeling of the detectors, a response function $R(f)$ is derived, which is then used to convert the digitized electrical output into strain $h(f)$ using the expression,
\begin{equation} \label{eq:voltage_to_strain}
    d(f) = \frac{1}{L} e(f) R(f)
\end{equation}
where e(f) is the digitized signals from the output photo-detectors, R(f) is the response function that converts e(f) into the differential displacement of the two arms of the detector and L is the average (macroscopic) length of the two arms. 

The response function of a gravitational wave detector, in the frequency domain, can be written as \cite{Sun_2020},
\begin{equation}
\label{eq:response_function}
    R(f)=\frac{1+A(f)D(f)C(f)}{C(f)}
\end{equation}
where $C(f)$ is the sensing function corresponding to the response of the detector to differential changes in its two arms without any feedback control, $A(f)$ is the actuation function used to control the positions of the mirrors and $D(f)$ is any digital filter(s) used in the control loop. 

\subsection{Sensing function}\label{subsec:sensing}
\begin{figure}[htbp]
\centering
\includegraphics[width=0.95\columnwidth]{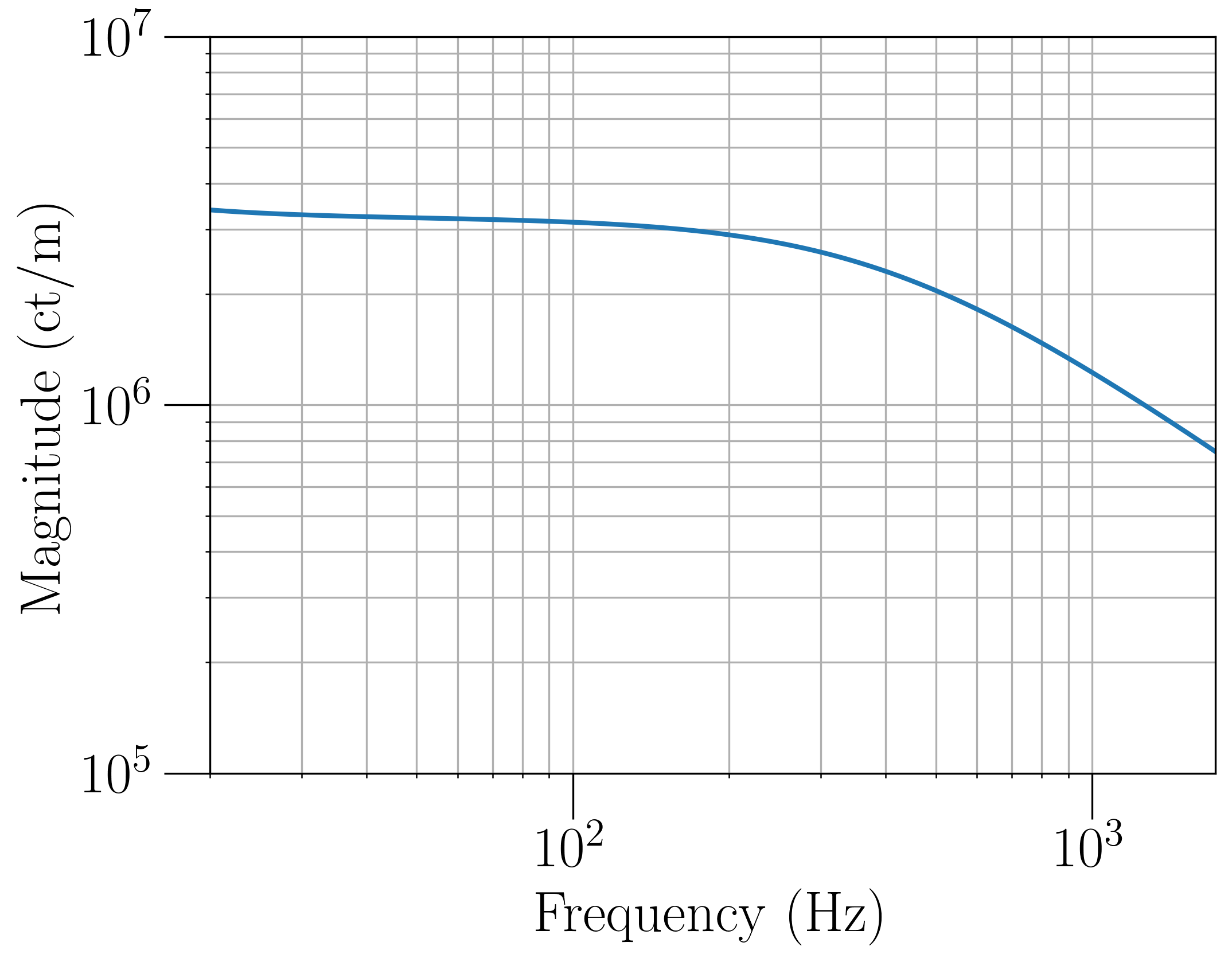}
\caption{The sensing function $C(f)$ used in our analysis. It is one of the sensing functions of the LIGO Hanford detector during the observing run O3 that is available in the {\tt pyDARM} package. The unit of $C(f)$ is the counts produced in the Analog-to-Digital converter at the output port for a meter differential length change in the two arms of the GW detector \cite{Sun_2020}.}
\label{fig:sensing_function}
\end{figure}

The sensing function $C(f)$ can be modeled in the frequency domain as \cite{calibLVK_O1O2,Sun_2020},
\begin{eqnarray}\label{eq:sensing_function}
    C(f) &=& \left(\frac{\kappa_C H_C}{1+i f f_{cc}^{-1}} \right)\left(\frac{f^2}{f^2+f_s^2-iff_sQ^{-1}} \right) \nonumber \\
      & & \ \times \ C_R(f)
\end{eqnarray}
where optical gain $H_C$ represents the overall gain, coupled-cavity pole frequency $f_{cc}$ defines the detector bandwidth, $f_s$ and $Q$ correspond to optical anti-spring pole frequency and its quality factor, respectively. The term $C_R$ represents the frequency dependencies not captured by the other terms (for example, the response of the electronics chain used for the digitization, etc.), and $\kappa_C$ is a scale factor representing the changes in the sensing function with respect to a reference time. The sensing function we use in our analysis is shown in Fig.~\ref{fig:sensing_function}. We use the {\tt pyDARM} package \cite{pydarm} to generate the calibration model used in this work. 
For LIGO detectors, during the past observing runs and for frequencies $\gtrsim 20$ Hz, the optical spring term (second term in Eq.~\ref{eq:sensing_function}) was usually close to one (for example, see \cite{lho_log_fs_Q, llo_log_fs_Q}).
Since in our work, we use $20 -1726$ Hz band as done in LVK analyses \cite{SGWBLVK_O1, SGWBLVK_O2, SGWBLVK_O3}, we treat the optical spring term in Eq.~\ref{eq:sensing_function} as constant and do not study its effects in this work. 

\subsection{Actuation function}\label{subsec:actuation}
\begin{figure}[htbp]
\centering
\includegraphics[width=0.95\columnwidth]{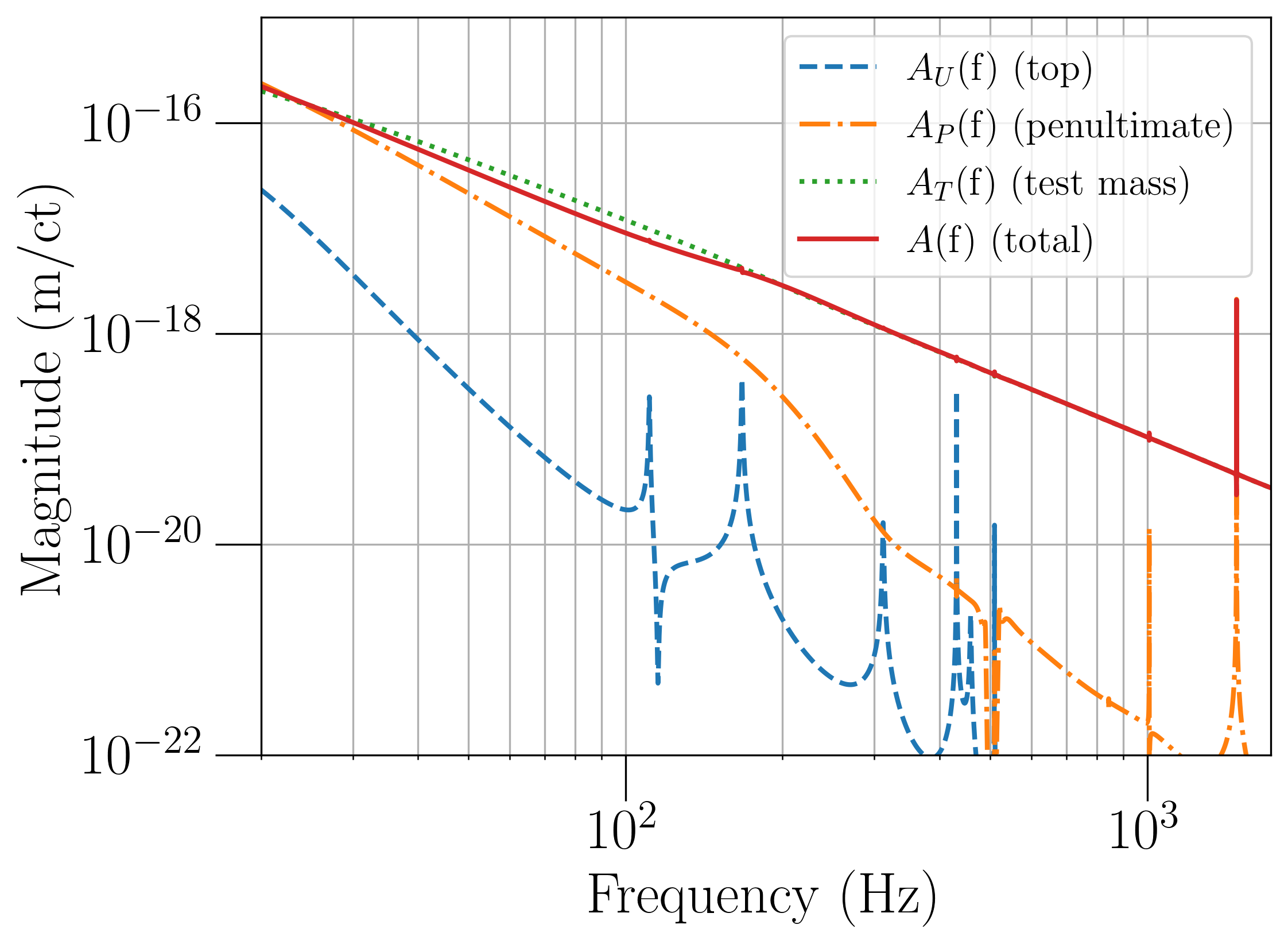}
\caption{The actuation functions of the bottom three stages (top, penultimate, and test mass stages) and the combined actuation function used in our analysis. This is one of the models of LIGO Hanford's main optic suspension during the observing run O3 available in the {\tt pyDARM} package. The unit of $A(f)$ is the differential length change produced in the two arms for a unit count in the Digital-to-Analog converter that drives the actuators \cite{Sun_2020}.}
\label{fig:actuation_function}
\end{figure}

The actuation function is modeled in the frequency domain as \cite{calibLVK_O1O2,Sun_2020},
\begin{eqnarray}\label{eq:actuation_function}
        A(f) &=& \kappa_U A_U(f) + \kappa_P A_P(f)+ \kappa_T A_T(f)
\end{eqnarray}
where $U$, $P$, and $T$ represent the lowest three stages of suspensions (upper intermediate mass, penultimate, and test mass stages) used to suspend the main optics \cite{aLIGO_2015, Sun_2020}. $A_i(f)$ (where $i=U,P,T$) are frequency-dependent actuation models of the three stages of the suspensions, including digital filters in the control path and analog responses of the three stages of suspensions \cite{Sun_2020}. The scale factors $\kappa_i$ capture any changes in the reference actuation model of each stage, and in general, they could be time- and frequency-dependent \cite{Tuyenbayev_2016}. The plots of actuation models for the three stages and the combined actuation model used in this work are shown in Fig.~\ref{fig:actuation_function}.

\subsection{Interferometer response function}
\begin{figure}[htbp]
    \includegraphics[width=0.95\columnwidth]{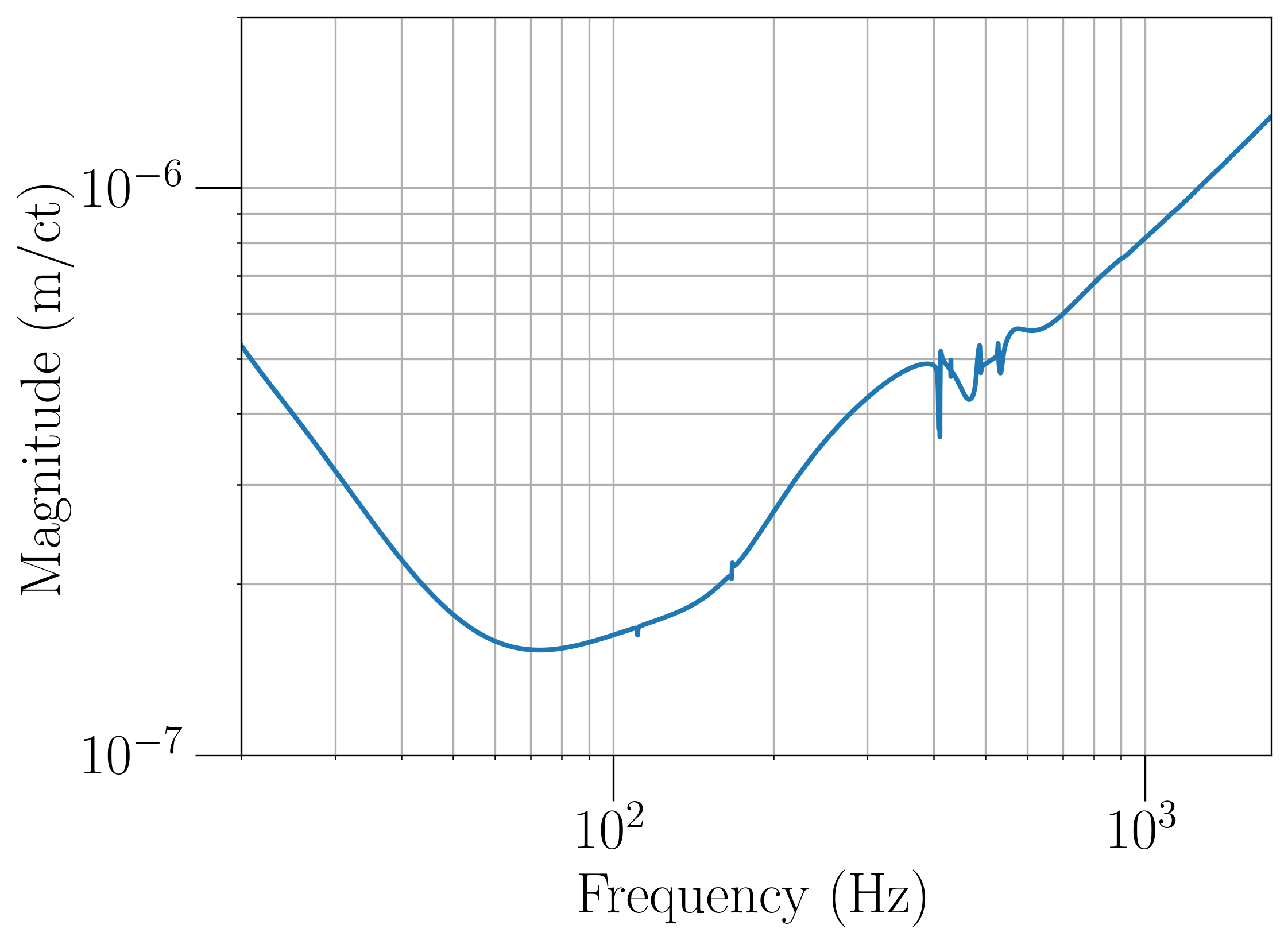}
    \caption{The reference response function $R(f)$ used in our analysis.}
    \label{fig:response_function}
\end{figure}

Apart from the notch filters used to prevent the excitation of resonances of the test mass suspensions, $D(f)$ is a smooth function of frequency that is decided by the feedback control morphology used. The total response function, as shown in Eq.~\ref{eq:response_function}, is a function of $C(f)$, $A(f)$, and $D(f)$. Fig.~\ref{fig:response_function} shows the response function we use in our analysis.

\section{Analysis Method}\label{sec:analysis_method}
In this work, we look at the effects of calibration uncertainties on the recovery of GWB and on the parameter estimation of the recovered GWB. Specifically, we look at the isotropic GWBs described by power-law models with power-law indices of $\alpha = {0, 2/3, 3}$ (see Sec.~\ref{sgwb_model}). 

If the response function used to calibrate the digitized signal in Eq.~\ref{eq:voltage_to_strain} is 
not the true response function, then we get,
\begin{eqnarray}\label{eq:d_true}
  d_{\rm true}(f) & = & d_{\rm calc}(f) \times \frac{R_{\rm true}(f)}{R_{\rm calc}(f)} \\
   & = & d_{\rm calc}(f) \times \Lambda(f)
\end{eqnarray}
where {\it true} and {\it calc} correspond to the true and calculated quantities respectively. In the above Eq.~\ref{eq:d_true}, we have defined $\Lambda(f)$ as,
\begin{equation}
    \Lambda(f) = \frac{R_{\rm true}(f)}{R_{\rm calc}(f)}
\end{equation}
for convenience. The uncertainties in the calibration process enter the GW analyses as $\Lambda(f)$ shown above. We note here that $R_{\rm true}(f)$, with measurement uncertainty, can be calculated using a length (or frequency) reference such as a photon calibrator \cite{glasgow_pcal, GEO_pcal, Virgo_pcal, LIGO_initial_pcal, pcal}, but due to difficulty in the implementation $R_{\rm calc}(f)$ is traditionally used in the calibration process leading to the difference we see in the Eq.\ref{eq:d_true}. The $R_{\rm true}(f)$ is usually in a non-parametric form while $R_{\rm calc}(f)$ is parameterized with a relatively small number of parameters (Eq.~\ref{eq:response_function}). Hence from an implementation point of view, $R_{\rm calc}(f)$ is more convenient. Because of the simple parameterization, changes in $R_{\rm calc}(f)$ can also be easily tracked, which is also important for calibration. Moreover, the ratios $\Lambda(f)$ are usually very close to one, and hence use of $R_{\rm calc}(f)$ is well justified. 

Due to the measurement uncertainties in $R_{\rm true}(f)$, the estimation of the ratios $\Lambda(f)$ has both systematic and statistical uncertainties associated with it. Using Eq.~\ref{eq:d_true} in Eqs.\ref{eq:Omega} and \ref{eq:Variance} we get,
\begin{equation} \label{eq:Omega_unc}
\hat{\Omega}_\alpha(f) = \frac{2}{T} \frac{\Re\left[d^*_{I,{\rm calc}}(f) d_{J,{\rm calc}}(f) \Lambda_I^*(f) \Lambda_J(f)\right]}{\gamma_{IJ}(f) S_\alpha(f)}    
\end{equation}
and
\begin{equation}\label{eq:Variance_unc}
   \sigma^2_{\hat{\Omega}_\alpha}(f) = \frac{1}{2 T \Delta f} \frac{P_{I,{\rm calc}}(f) P_{J,{\rm calc}}(f)}{\gamma^2_{IJ}(f) S^2_\alpha(f)} |\Lambda_I|^2  |\Lambda_J|^2 .
\end{equation}
The Eqs.~\ref{eq:Omega_unc} and \ref{eq:Variance_unc} provide a way to estimate the effects of calibration uncertainties on the signal estimate $\hat{\Omega}_\alpha$ and its variance $\sigma^2_{\hat{\Omega}_\alpha}$.
If we further assume that the ratios $\Lambda(f)$ are real, i.e., the difference is only in the magnitude, then we get,
\begin{eqnarray} 
  \label{eq:simplified_omega}
   \hat{\Omega}_\alpha(f) &=& \hat{\Omega}_{\alpha, {\rm nocal}}(f) \Lambda_I(f) \Lambda_J(f) \; ,\\   
   \label{eq:simplified_variance}
   \sigma^2_{\hat{\Omega}_\alpha}(f) &=&  \sigma^2_{\hat{\Omega}_\alpha, {\rm nocal}}(f) \Lambda_I^2(f) \Lambda_J^2(f) ,
\end{eqnarray}
where {\it nocal} subscript corresponds to the quantities calculated in the absence of calibration uncertainties that we want.  
With this assumption, the simulation becomes a little bit easier. We can start with $\hat{\Omega}_{\alpha, {\rm nocal}}(f)$ and $\sigma^2_{\hat{\Omega}_\alpha, {\rm nocal}}(f)$ calculated from the simulated data and using Eqs.~\ref{eq:simplified_omega}, \ref{eq:simplified_variance} and \ref{eq:combining_segments} we can estimate the effects of calibration uncertainties on the calculation of $\hat{\Omega}_\alpha(f)$ and $\sigma^2_{\hat{\Omega}_\alpha}(f)$. However, in Sec.\ref{sec:results} we also show the results without using this assumption.  
Since the response functions, $R_{I,J}$ themselves are functions of $A$ (Eq.~\ref{eq:actuation_function}), $C$ (Eq.~\ref{eq:sensing_function}) and $D$ the number of free parameters in the above equations becomes large. Due to the large number of parameters, it is difficult to calculate the effects analytically, so we use numerical simulation to calculate the effects. This method becomes more valuable when including a more complicated signal model and additional calibration parameters. 

For the results reported in this paper, we use one week of simulated data for Hanford and Livingston detectors using advanced LIGO design sensitivity \cite{prospects_LRR}. Here, one week of data is chosen to represent the traditional long-duration analyses of GWB and to avoid complexities arising from large SNRs in individual segments \cite{AllenRomano_1999}. We use publicly available LVK code packages \cite{stochastic_pipeline} to calculate $\hat{\Omega}_\alpha (t;f)$ and $\sigma_{\hat{\Omega}_\alpha} (t;f)$. We use standard search parameters of 192-sec segment duration and frequencies from 20 Hz to 1726 Hz with a frequency resolution of 1/32 Hz as used in the LVK isotropic GWB searches \cite{SGWBLVK_O1, SGWBLVK_O2, SGWBLVK_O3}. In this work, we use the same calibration model for Hanford and Livingston detectors described in Sec.~\ref{sec:calibration_model}. 

We do the following to calculate the effects of calibration uncertainties on the recovery of GWB signal. As indicated in the Eqs.~\ref{eq:simplified_omega} and \ref{eq:simplified_variance}, we multiply the $\hat{\Omega}_{\alpha, {\rm nocal}}(t; f)$ and $\sigma^2_{\hat{\Omega}_\alpha, {\rm nocal}}(t; f)$ estimators of each segment calculated using LVK code packages by distributions representing the ratios $\Lambda(f)$. We assume Gaussian distributions for $\Lambda(f)$, centered at one with standard deviations defined by the desired calibration uncertainty. We also truncate the Gaussian distribution at 2-sigma points on both sides to avoid the realization of unrealistic values for $\Lambda(f)$ (for example, values close to zero or even negative). Then, using Eqs.~\ref{eq:combining_segments}, we combine the segment-wise and frequency-dependent results of $\hat{\Omega}_\alpha(t;f)$ $\sigma_{\hat{\Omega}_\alpha}(t;f)$ to get the final estimate and its uncertainty.
Then we use SNR, defined in a frequentist approach \cite{Freq_Baysian}, given by, 
\[{\rm SNR} = \frac{\hat{\Omega}_\alpha}{\sigma_{\hat{\Omega}_\alpha}}\]
as the detection statistics in the search for an isotropic GWB. We then compare these results against the results obtained without any calibration uncertainties. Since the difference between these results is just the application of calibration uncertainties, the differences would typically show the effects of calibration uncertainties on $\hat{\Omega}_\alpha$ and $\sigma^2_{\hat{\Omega}_\alpha}$.   

We further look at the effects of calibration uncertainties on the parameter estimation, specifically on the $\hat{\Omega}_\alpha$ and $\hat{\alpha}$, by varying the values of various parameters in the $R(f)$ (see Eqs.\ref{eq:response_function}, \ref{eq:sensing_function}. \ref{eq:actuation_function}).

\section{results}\label{sec:results}
In this section, we present the results of our studies. To generate these results, we initially assume that the ratios of response function $\Lambda(f)$ are real and hence use Eqs. \ref{eq:simplified_omega} and \ref{eq:simplified_variance}. We note that this assumption is used to marginalize calibration uncertainties in the LVK isotropic GWB analyses \cite{SGWBLVK_O1, SGWBLVK_O2, SGWBLVK_O3}. However, for comparison, we also produce results by additionally using 1-sigma phase uncertainties of $5 ^{\circ}$, the maximum of what was seen in LIGO detectors during the observing run O3 \cite{Sun_2020}. This is to show how much phase uncertainties that are currently not included in the GWB analyses affect the final results. At each frequency, we model the magnitude of $\Lambda(f)$ by a Gaussian distribution with a mean one and standard deviation $\sigma_{\Lambda(f)}$ that is small compared to one and phase of $\Lambda(f)$ by a Gaussian distribution with a mean zero and standard deviation of $5 ^{\circ}$. As indicated earlier, we also truncate the Gaussian distribution at 2-sigma values to avoid unrealistic realizations of $\Lambda(f)$.

\subsection{Effect of calibration uncertainties on the isotropic GWB detection} \label{subsec:effect_detection}
\begin{figure*}
  \centering
  \begin{tabular}{ccc}
    $\alpha = 0$ & $\alpha = 2/3$ & $\alpha = 3$ \\
    \includegraphics[width=0.33\textwidth]{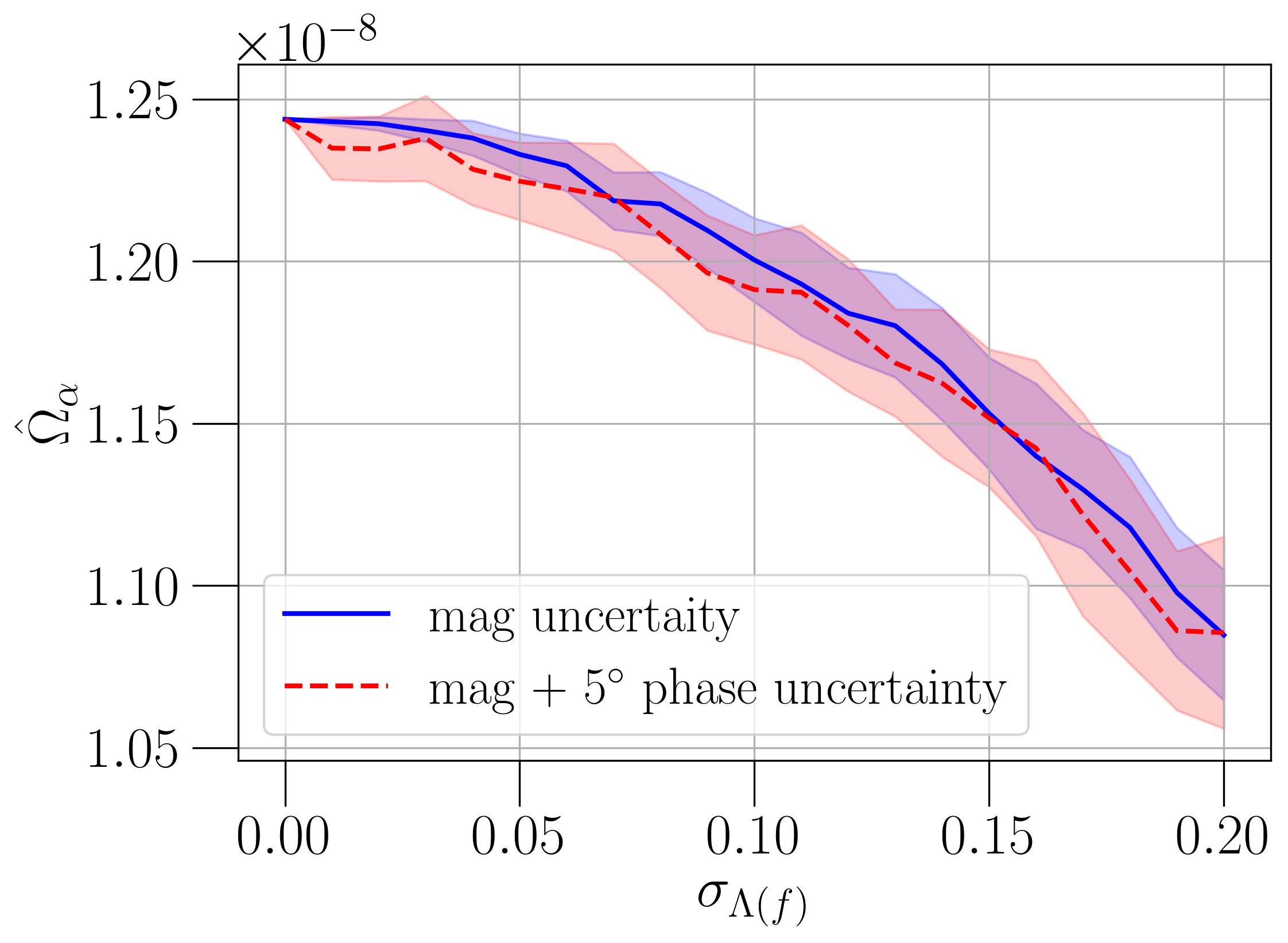} &
    \includegraphics[width=0.33\textwidth]{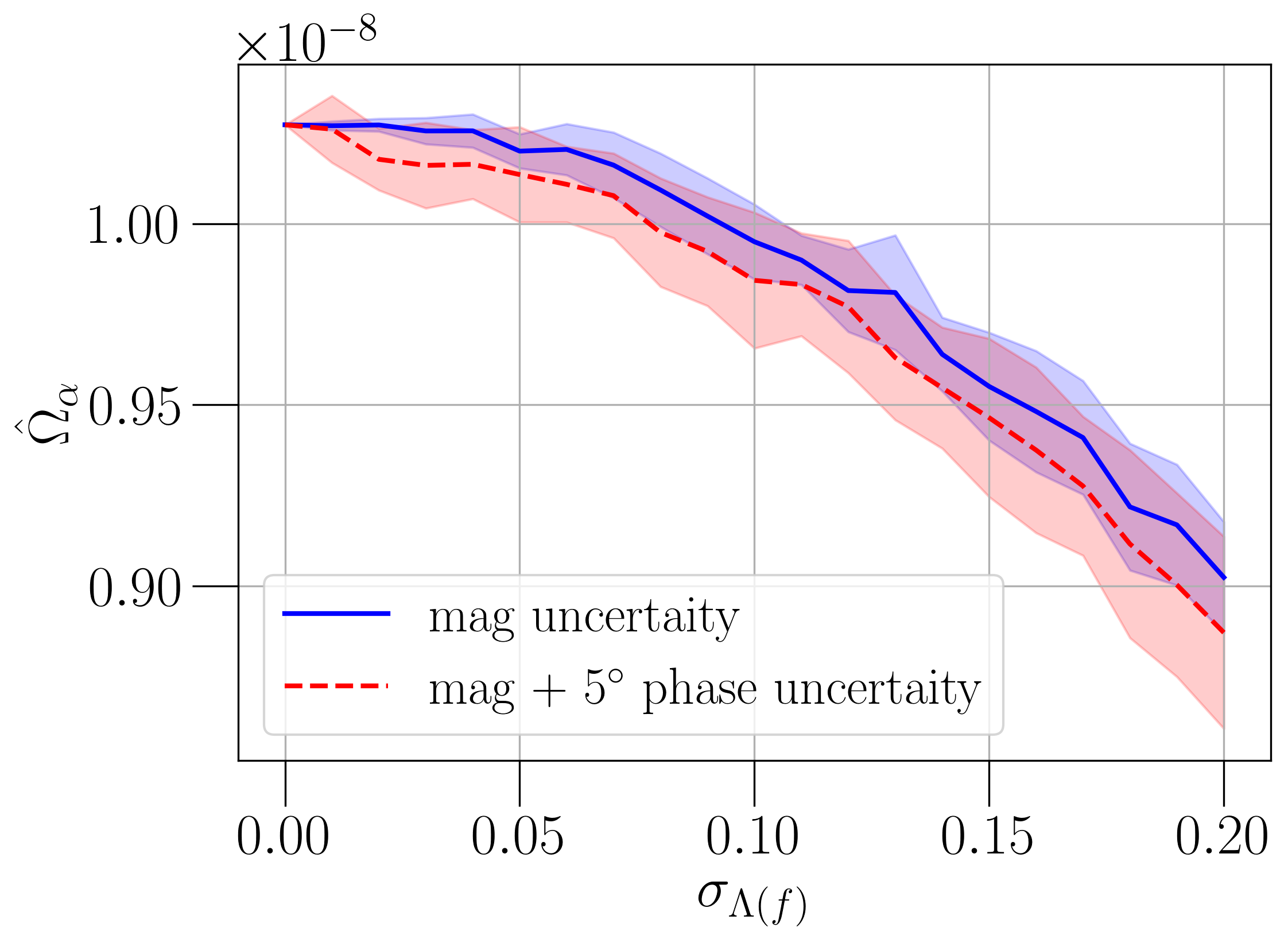} &
    \includegraphics[width=0.33\textwidth]{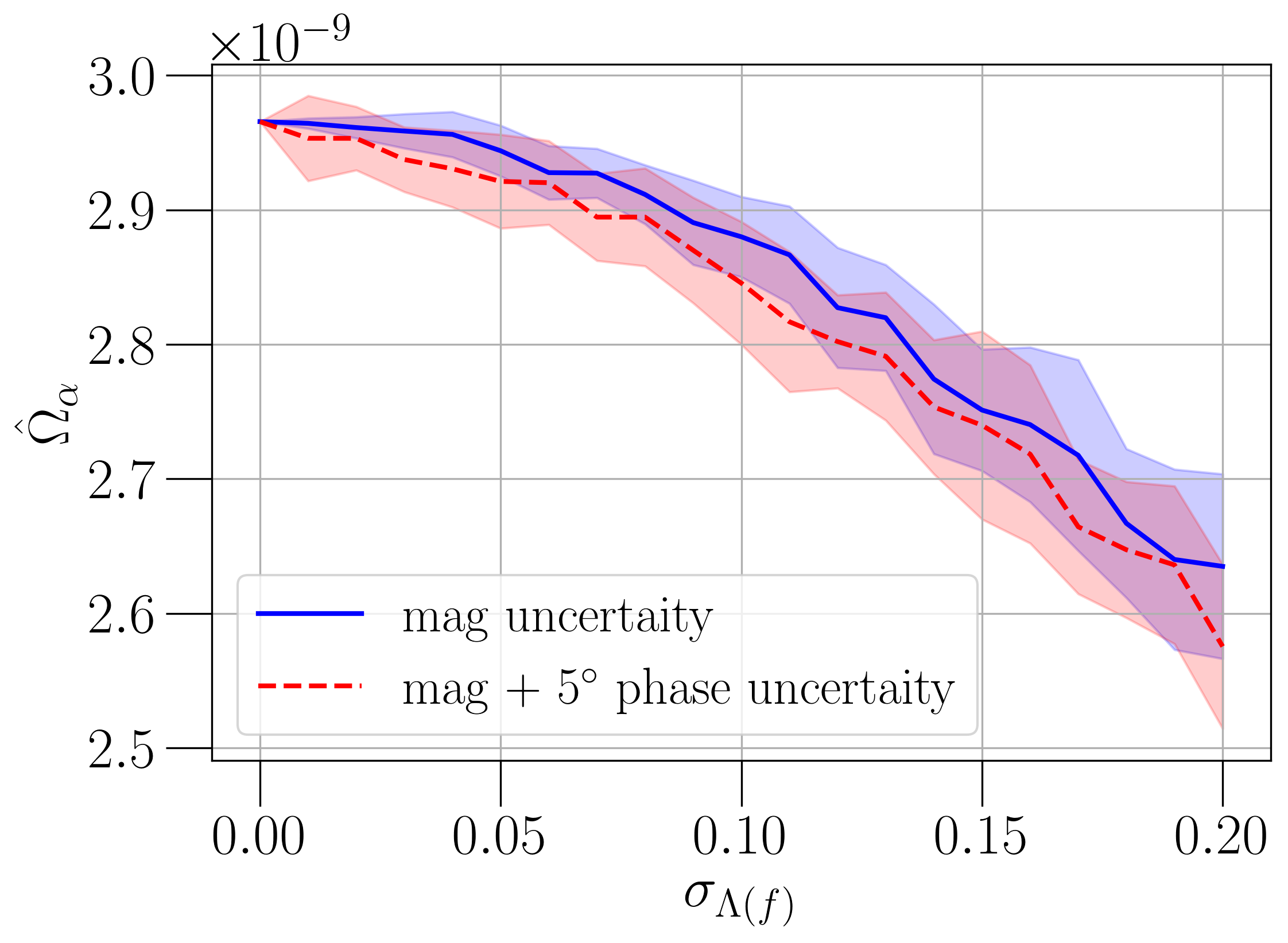} \\
    \includegraphics[width=0.3\textwidth]{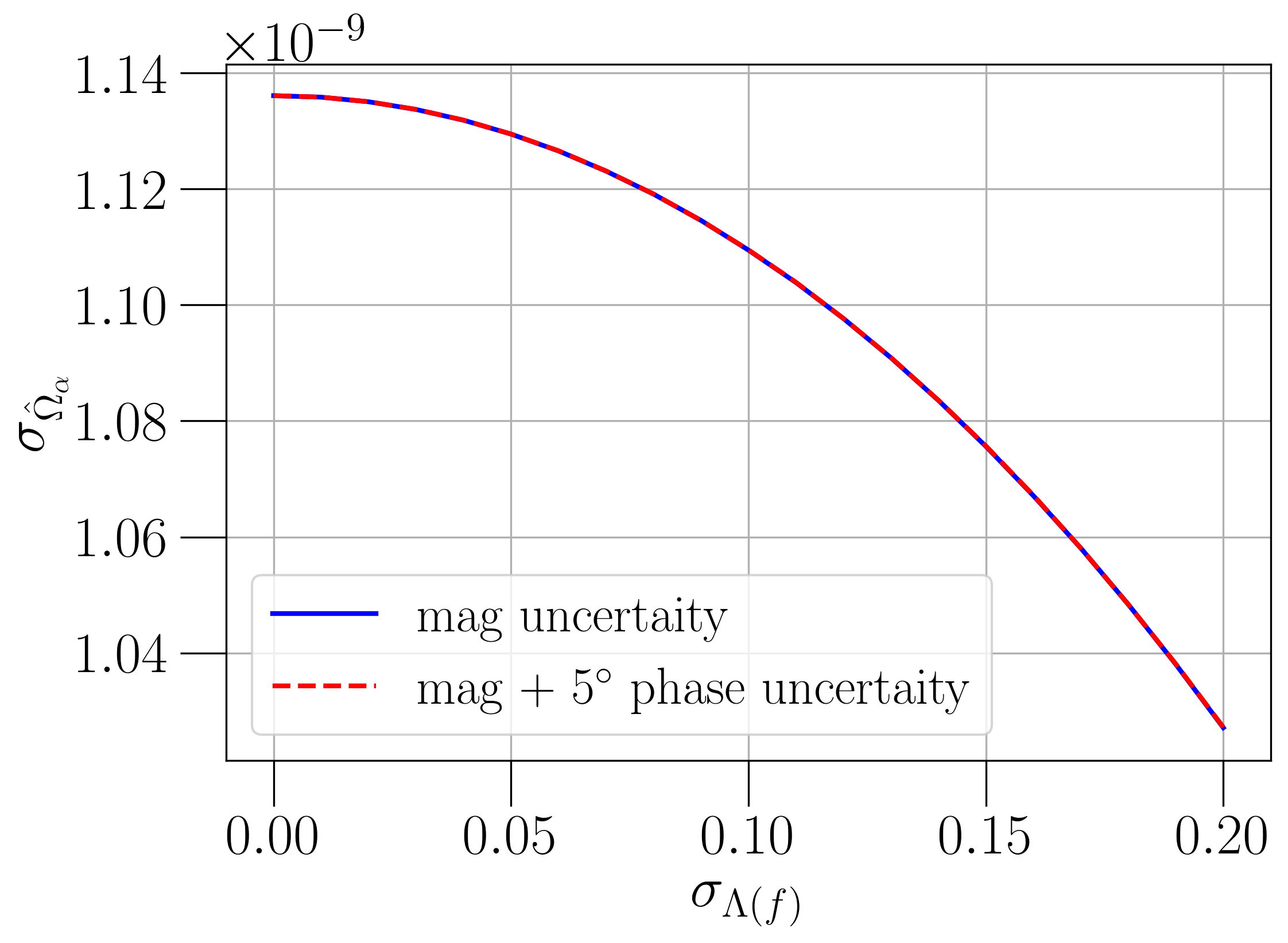} &
    \includegraphics[width=0.3\textwidth]{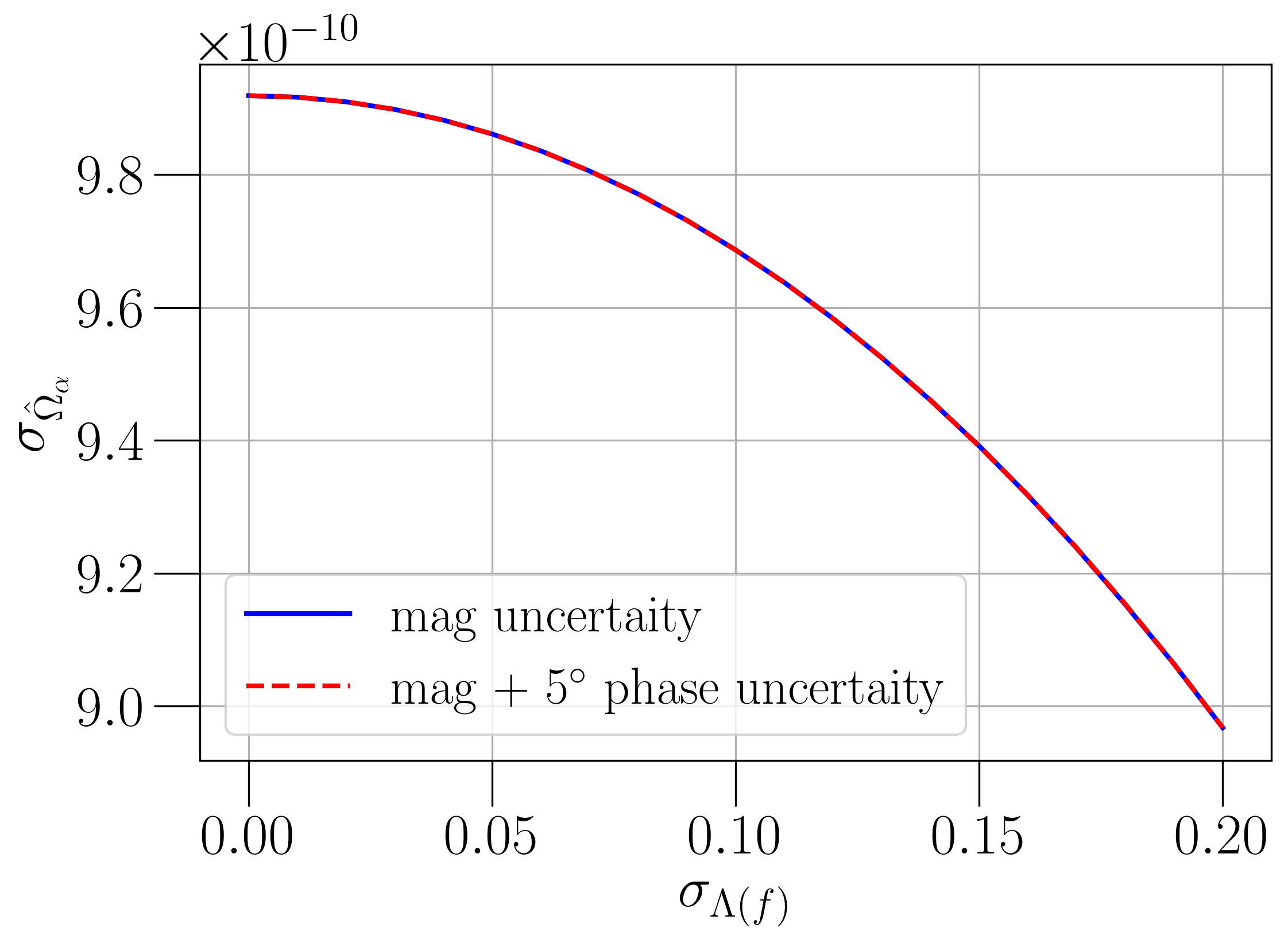} &
    \includegraphics[width=0.3\textwidth]{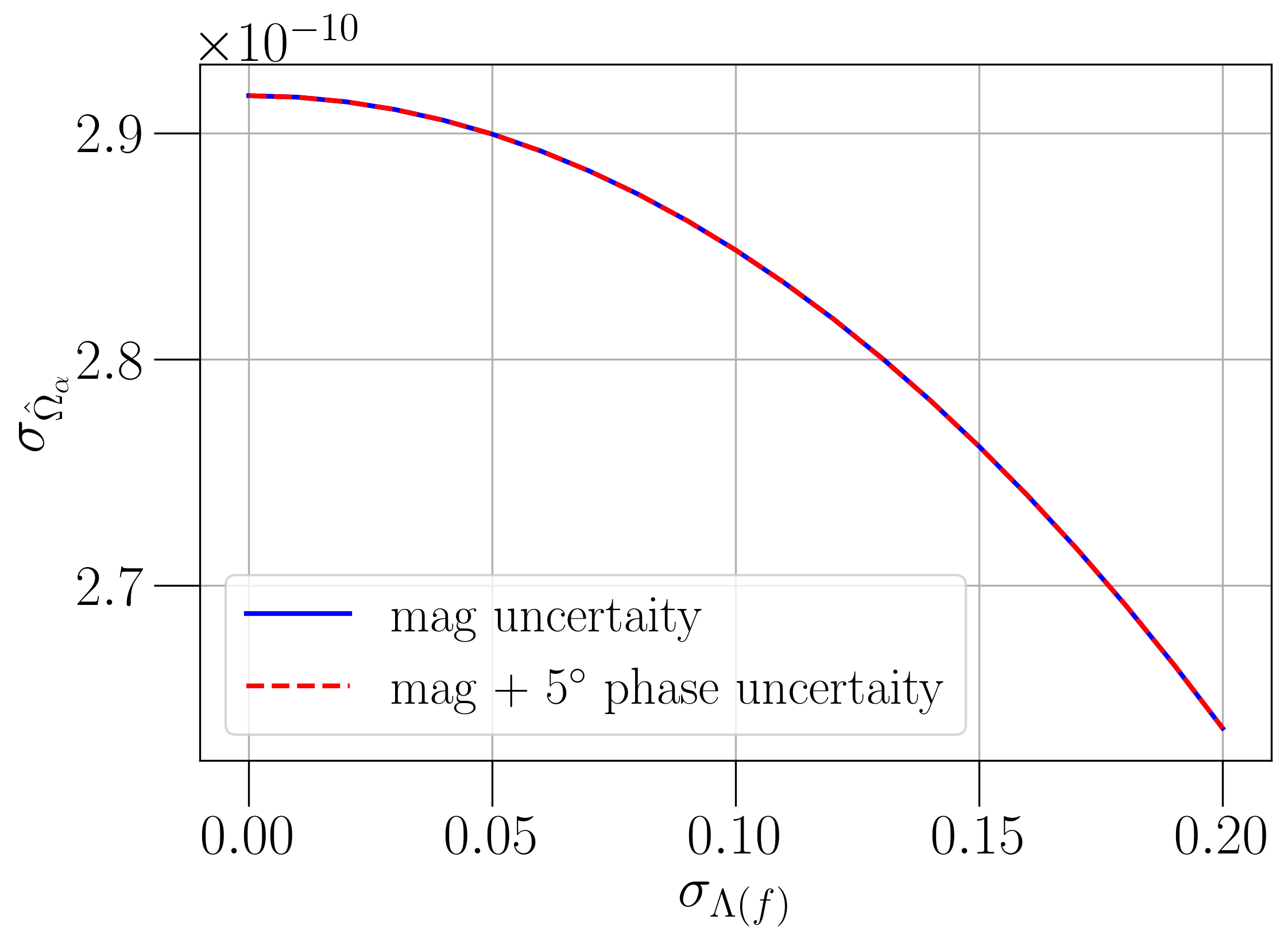} \\
    \includegraphics[width=0.3\textwidth]{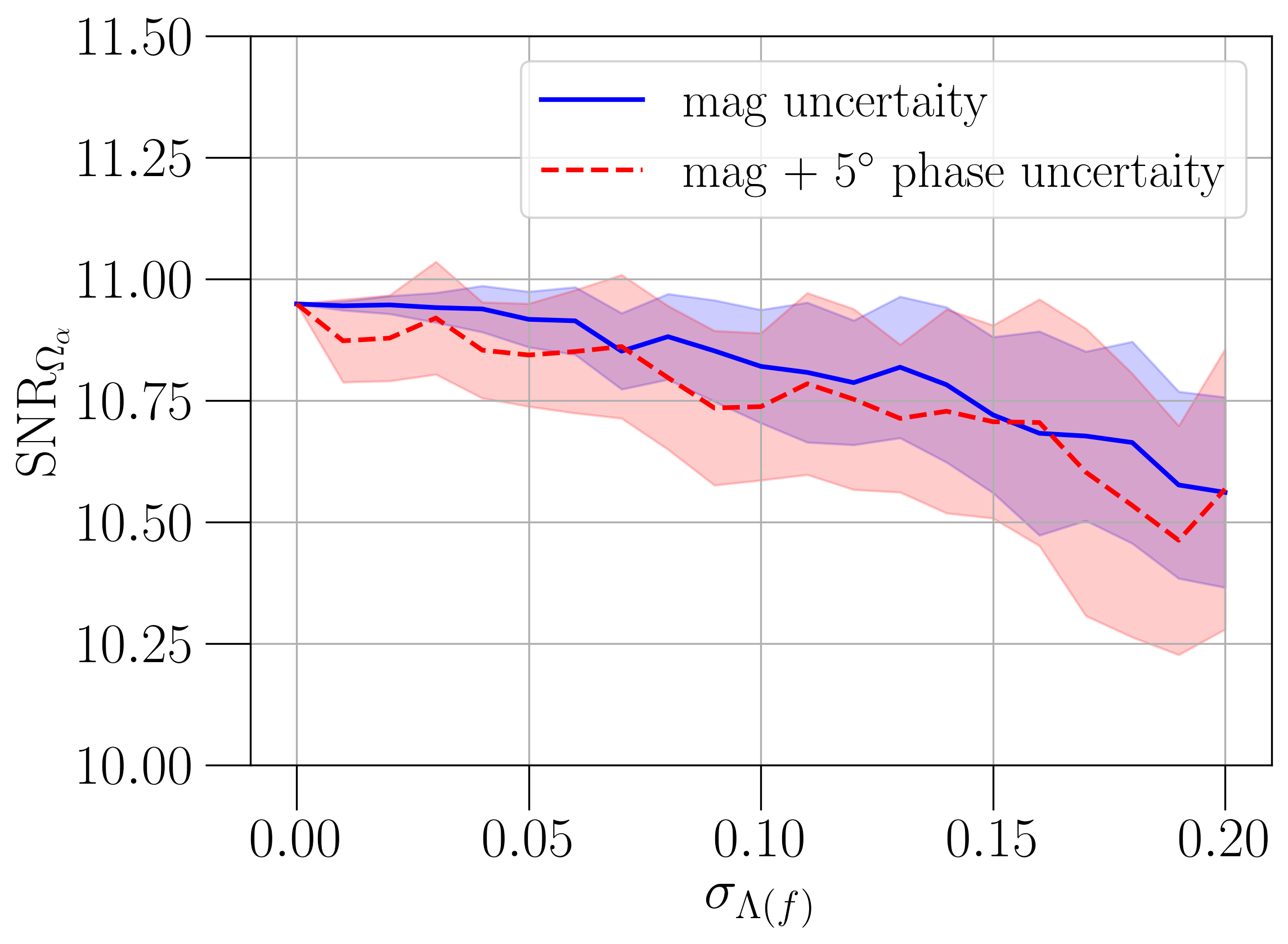} &
    \includegraphics[width=0.3\textwidth]{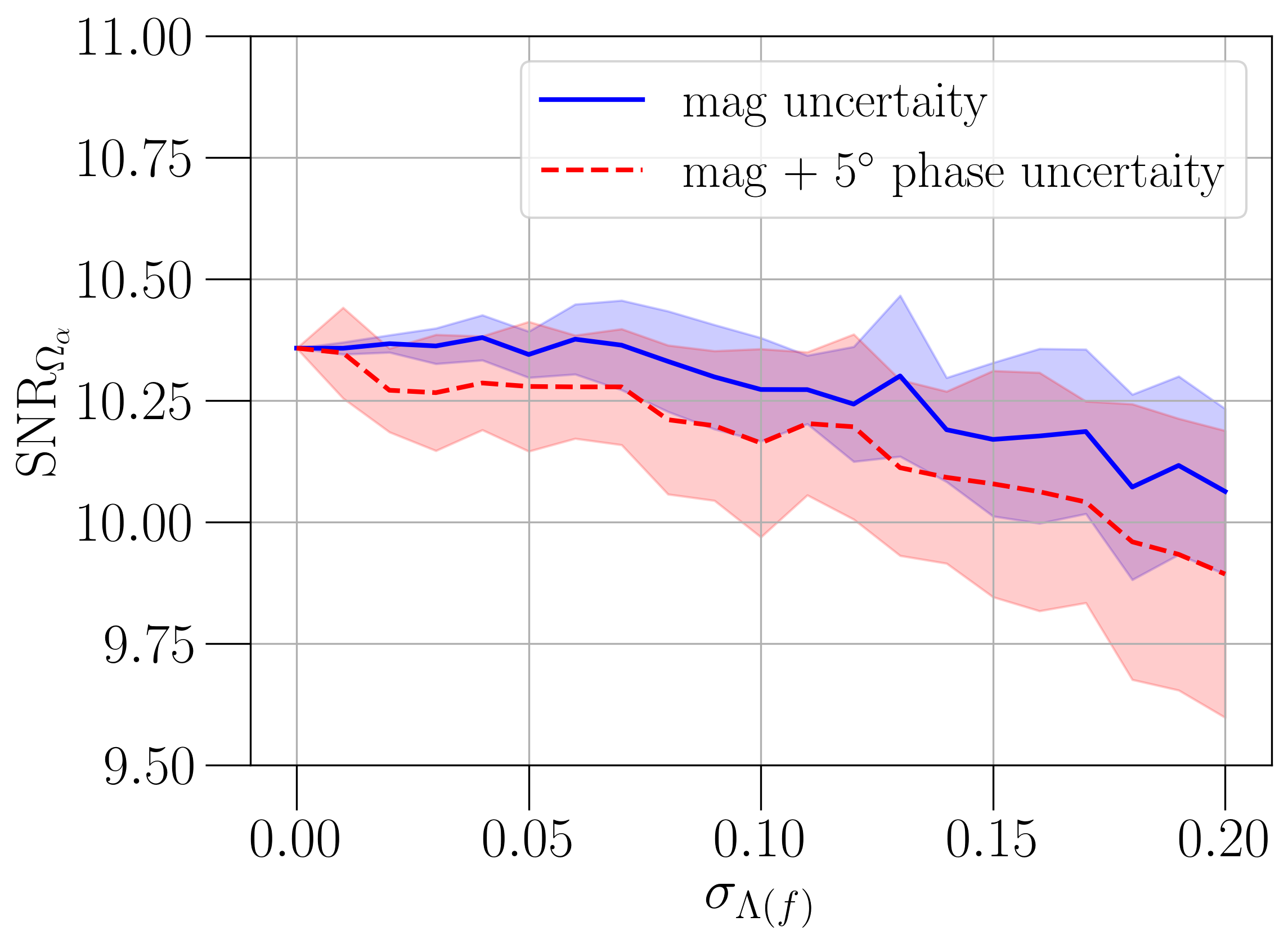} &
    \includegraphics[width=0.3\textwidth]{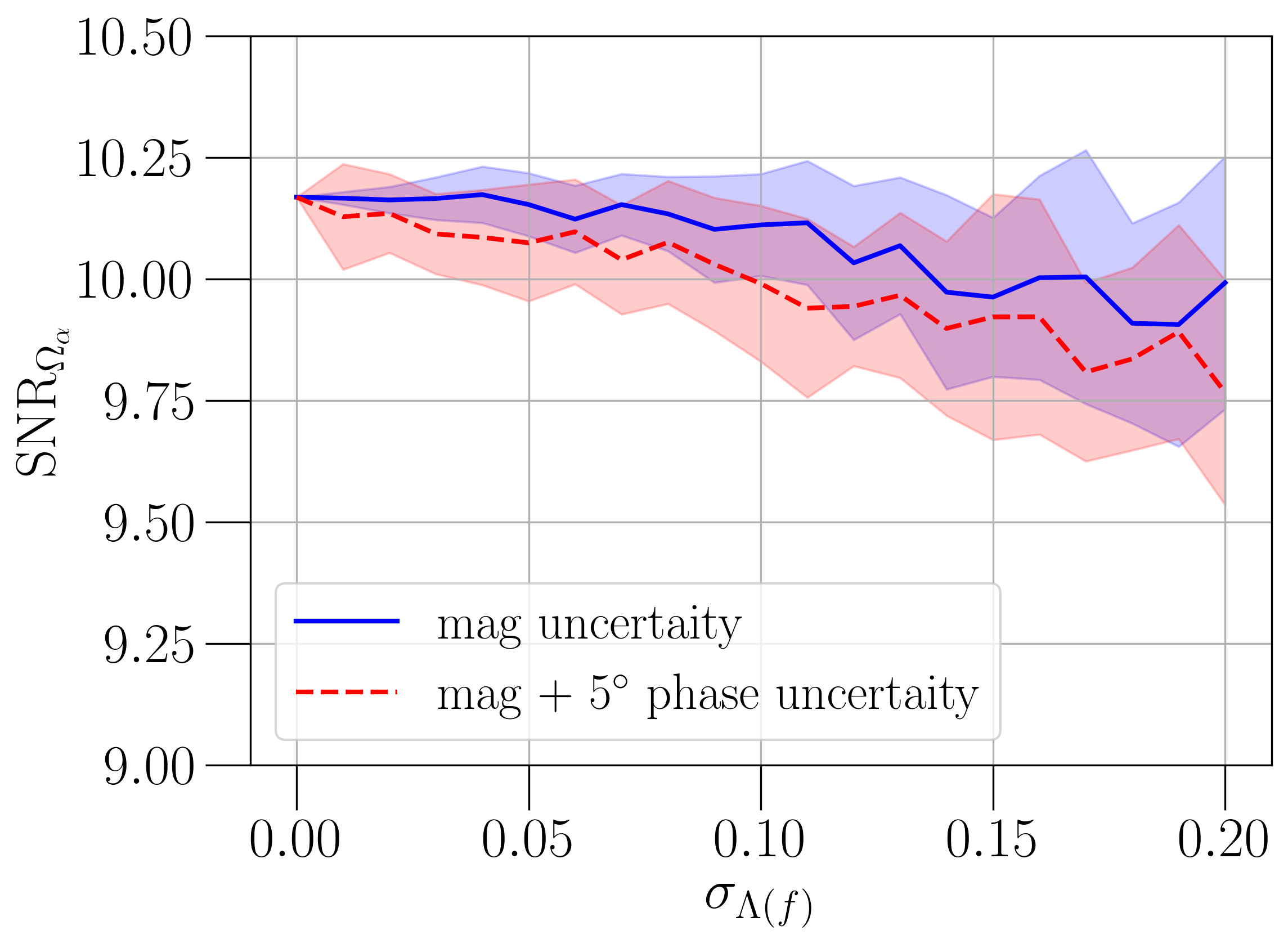}
  \end{tabular}
    \caption{Plots showing the effect of calibration uncertainty on the recovery of $\Omega_{\alpha}$, $\sigma_{\hat{\Omega}_\alpha}$ and SNR for injected isotropic GWB signals described by $\alpha={0,2/3,3}$. The calibration uncertainty is quantified by the standard deviation of the Gaussian distribution $\sigma_{\Lambda(f)}$ used for the different realizations of $\Lambda(f)$. The solid (blue) line corresponds to no phase uncertainty, while the dotted (red) line corresponds to $5 ^{\circ}$  1-sigma phase uncertainty.}
    \label{fig:all_detection_plots}
\end{figure*}

The recovered values of the $\hat{\Omega}_{\alpha}$, $\sigma_{\hat{\Omega}_\alpha}$ and SNR at various levels of calibration uncertainties for the three power law models $\alpha={0, 2/3, 3}$ are shown in Fig.~\ref{fig:all_detection_plots}. In this analysis, we increase the uncertainty from 
$0 \, \%$ to $20 \, \%$ in steps of $2 \, \%$. We also repeat the analysis $20$ times, regenerating the $\Lambda(f)$ values $20$ times at each uncertainty level to calculate the spread on the recovered values. We also compare the results, including 1-sigma phase uncertainties of $5 ^{\circ}$.

From the plots, we see that as we increase the values of uncertainties, there are changes in the recovered values of $\hat{\Omega}_{\alpha}$, $\sigma_{\hat{\Omega}_\alpha}$, and SNR. The recovered values are underestimated, and the trends are similar for the three $\alpha$ values. However, the changes in the recovered SNRs are small, almost negligible, below the calibration uncertainties of $\sim 10 \%$. Since SNR is generally used as a detection statistic, this suggests that the detection of an isotropic GWB is not significantly affected by the uncertainties in the calibration. We also see a slight reduction in the SNR for larger calibration uncertainties. The SNR dependence on the calibration uncertainty goes as $(1- \sigma_{\Lambda (f)}^2)$ where $\sigma_{\Lambda(f)}$ is the standard deviation of the Gaussian distribution used for the different realizations of $\Lambda(f)$. This quadratic dependence agrees with the results previously reported in the literature \cite{allen_cal}. 

The $\hat{\Omega}_{\alpha}$, $\sigma_{\hat{\Omega}_\alpha}$ change by $\sim 10 \%$ when we change the uncertainty of response function by $\sim 20 \%$. The reduction in the estimated $\sigma_{\hat{\Omega}_\alpha}$ can be attributed to how we combine different time segments and frequency bins. Since we use weighted average method (see Eq.~\ref{eq:combining_segments}), any downward fluctuations in individual $\sigma_{\hat{\Omega}_\alpha}(t; f)$ due to calibration uncertainties will bring down the final $\sigma_{\hat{\Omega}_\alpha}$. A similar effect could be attributed to the reduction in the final $\Omega_{\alpha}$. This suggests that the recovered values of $\Omega_{\alpha}$ and $\sigma_{{\Omega}_\alpha}$ are biased in the presence of calibration uncertainties. Since the upper limits on $\Omega_{\alpha}$, for example, 95 \% upper limit in the frequentist approach, can be written as
\[ \Omega_{\alpha , 95 \%} \approx \hat{\Omega}_{\alpha} + 2 \, \sigma_{\hat{\Omega}_\alpha} , \]
calibration uncertainties are also expected to bias the upper limit calculations. From our results, we see that if the calibration (magnitude) uncertainty is $10 \% \, (\sigma_{\Lambda(f)} = 0.1)$, the upper limit would be underestimated by $\sim 3 \%$. Since this dependence on the calibration uncertainty is quadratic, this effect could become significant at larger calibration uncertainties. Such biases are not completely taken into account when estimating $\Omega_{\alpha}$ or while calculating upper limits on $\Omega_{\alpha}$ in the analyses reported in the literature \cite{SGWBLVK_O1, SGWBLVK_O2, SGWBLVK_O3} and need to be accounted for in future analyses. The plots also suggest that including phase uncertainties at the level of $\lesssim 5 ^{\circ}$ does not change the results significantly. Hence, as done in LVK analyses \cite{SGWBLVK_O1, SGWBLVK_O2, SGWBLVK_O3}, 
phase uncertainties can be neglected if they are $\lesssim 5 ^{\circ}$ when searching for isotropic GWB using LVK data. 

\subsection{Effects of the calibration uncertainties on the parameter estimation of isotropic GWBs}
The second part of the study looks at the effects of calibration uncertainties on estimating the parameters of the isotropic GWB signals. Here we mainly focus on the estimation of $\Omega_{\alpha}$ and $\alpha$ (see Eq.~\ref{eq:power_law}). In Sec.~\ref{subsec:effect_detection}, Fig.~\ref{fig:all_detection_plots} already shows the effect of the uncertainties of the response function as a whole on the recovery of $\Omega_{\alpha}$. Instead of the uncertainties of the total response function, in this section, we look at the effects of individual calibration parameters on the recoveries of $\Omega_{\alpha}$ and $\alpha$. Since we are using the parameters that make up the calibration model, in the literature, this is considered a physically motivated approach to include calibration uncertainties in the signal analyses \cite{Ethan20, Salvo21}. In this study we mainly focus on the parameters $\kappa_C$, $f_{cc}$ (see Sec.~\ref{subsec:sensing}), $\kappa_U$, $\kappa_P$ and $\kappa_T$ (see Sec.~\ref{subsec:actuation}). Other parameters in the response function tend to be more or less constant during an observing run, or their effects are small, and hence we do not include them here. 

\begin{figure*}
  \centering
  \begin{tabular}{ccc}
    $\alpha = 0$ & $\alpha = 2/3$ & $\alpha = 3$ \\
    \includegraphics[width=0.33\textwidth]{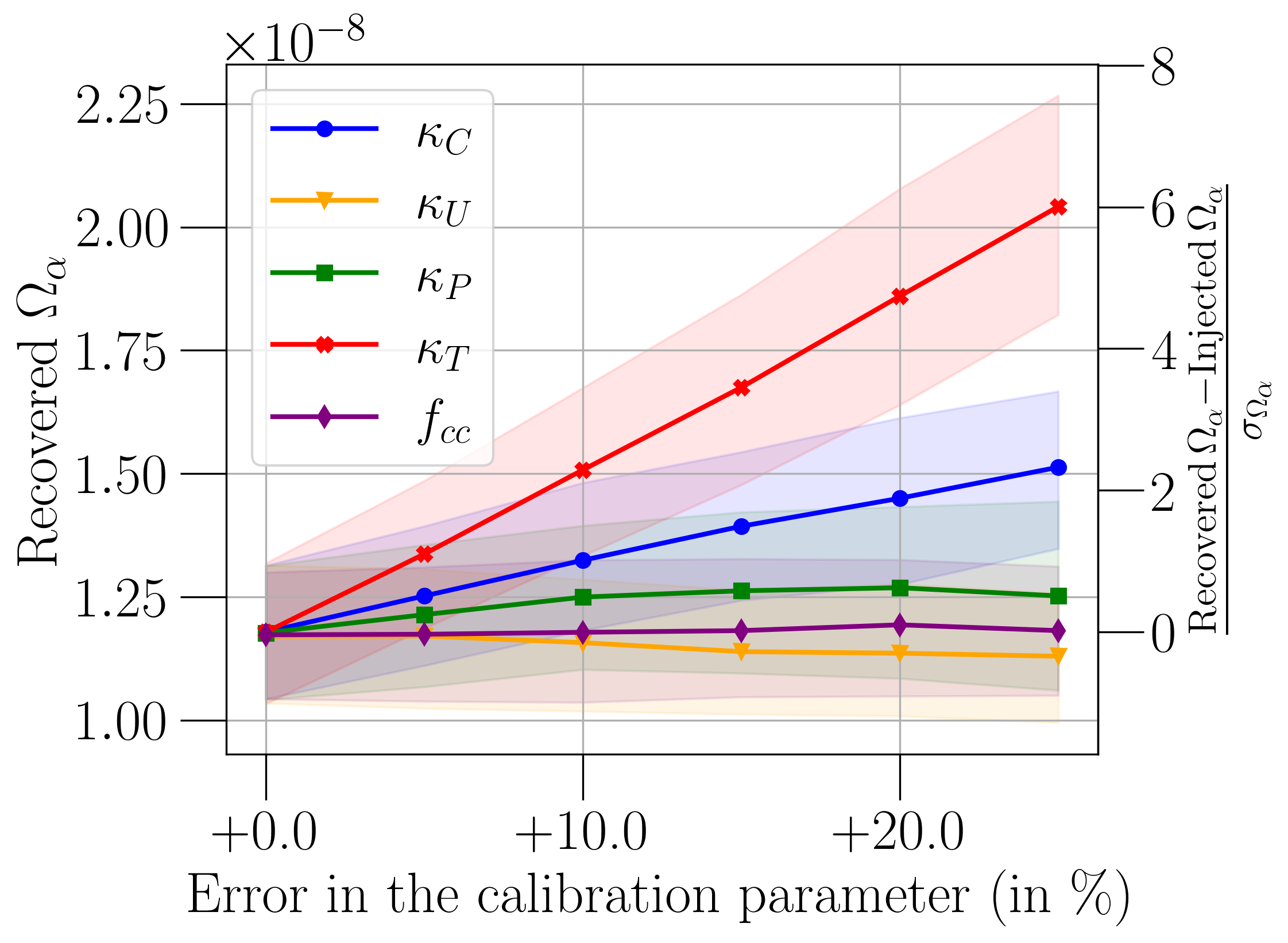} &
    \includegraphics[width=0.33\textwidth]{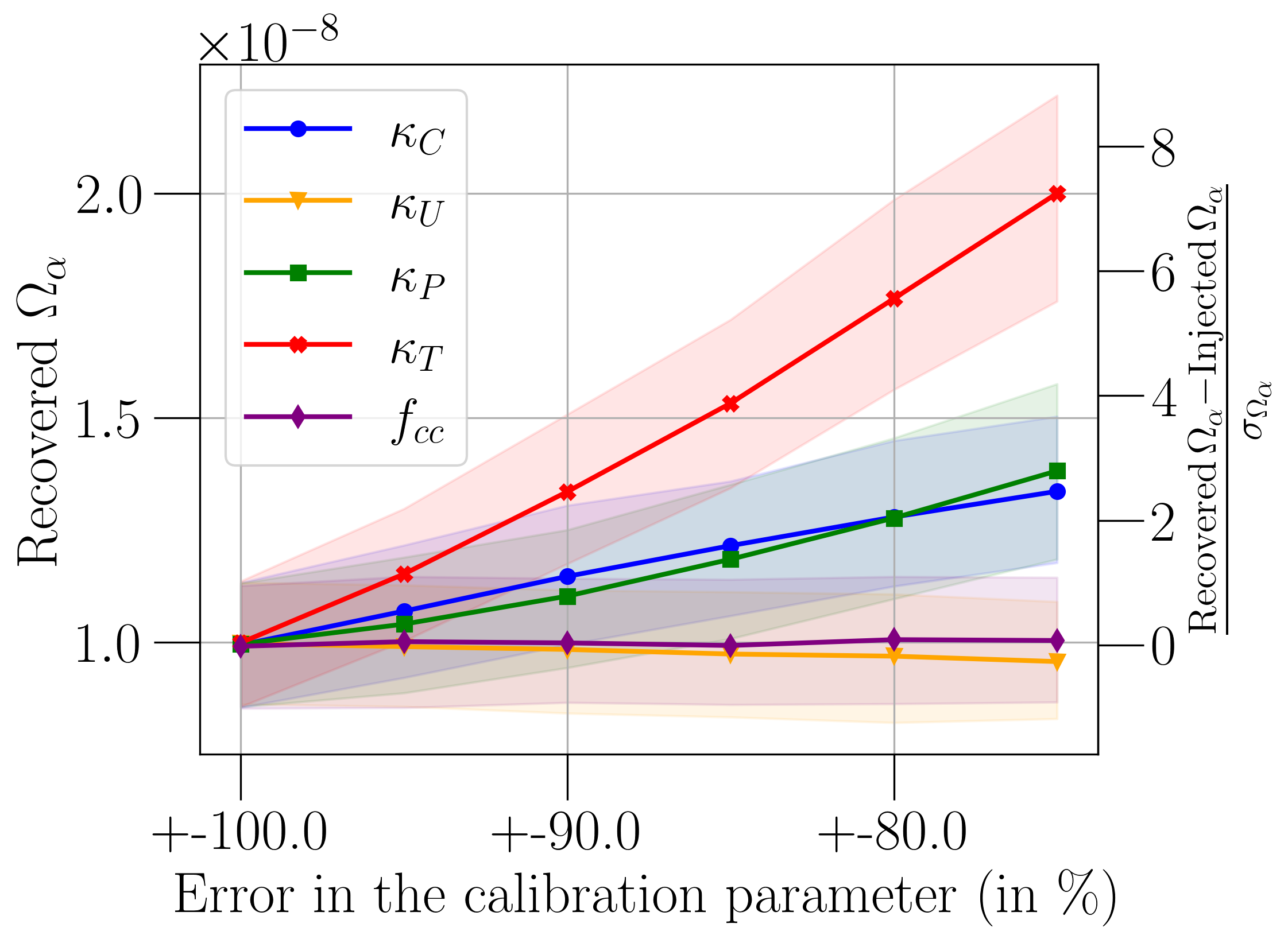} &
    \includegraphics[width=0.33\textwidth]{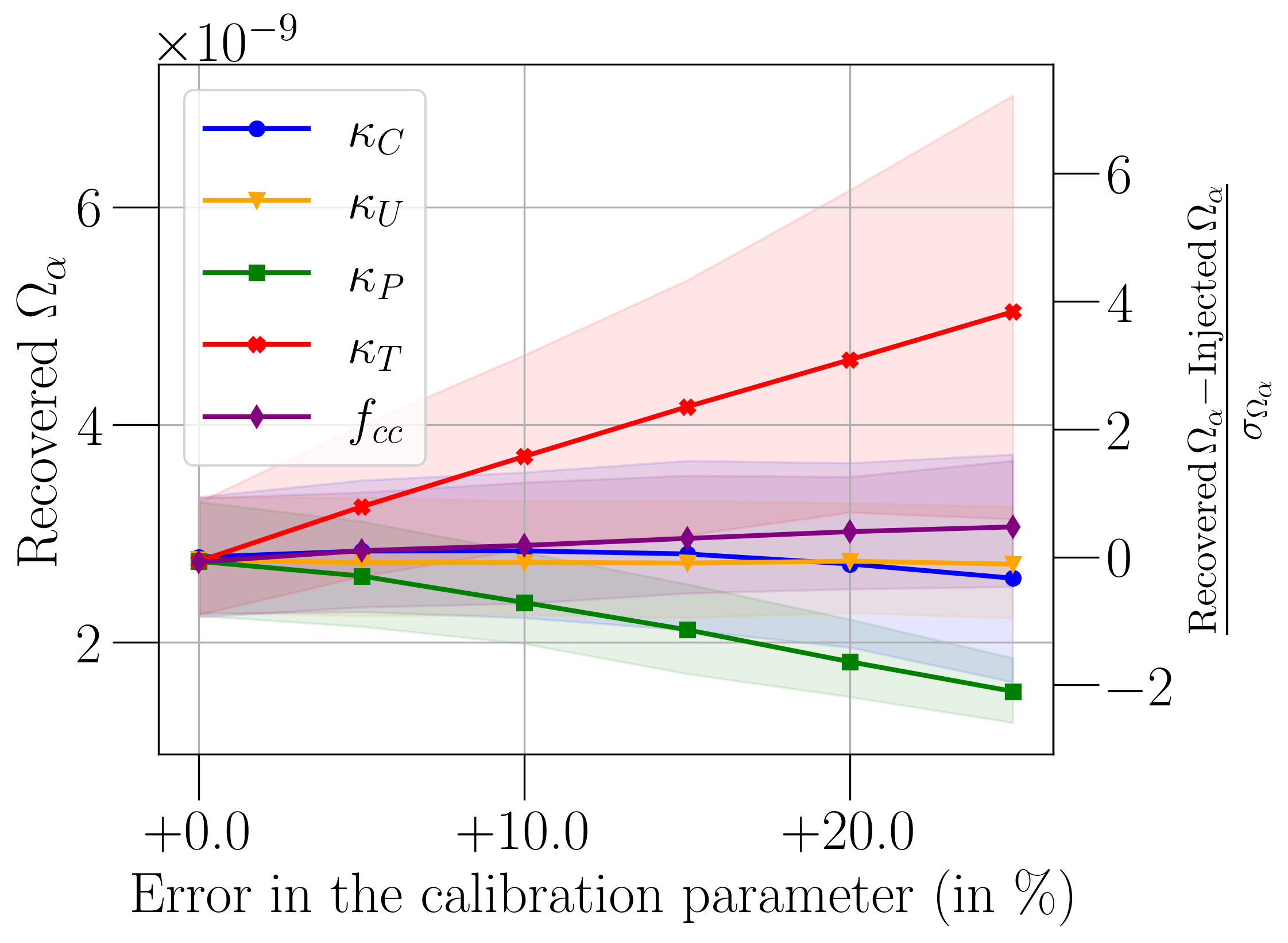} \\
    \includegraphics[width=0.3\textwidth]{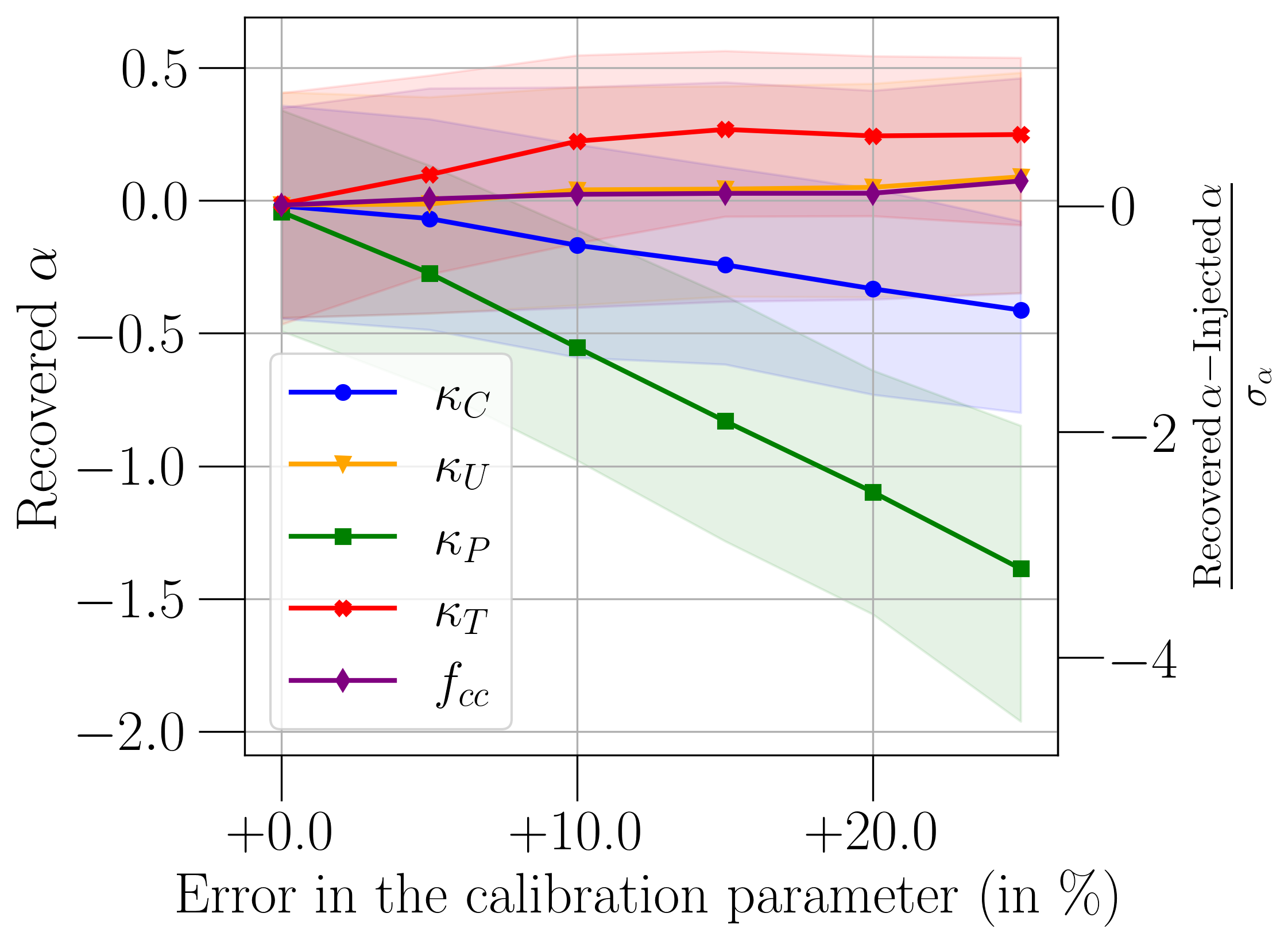} &
    \includegraphics[width=0.3\textwidth]{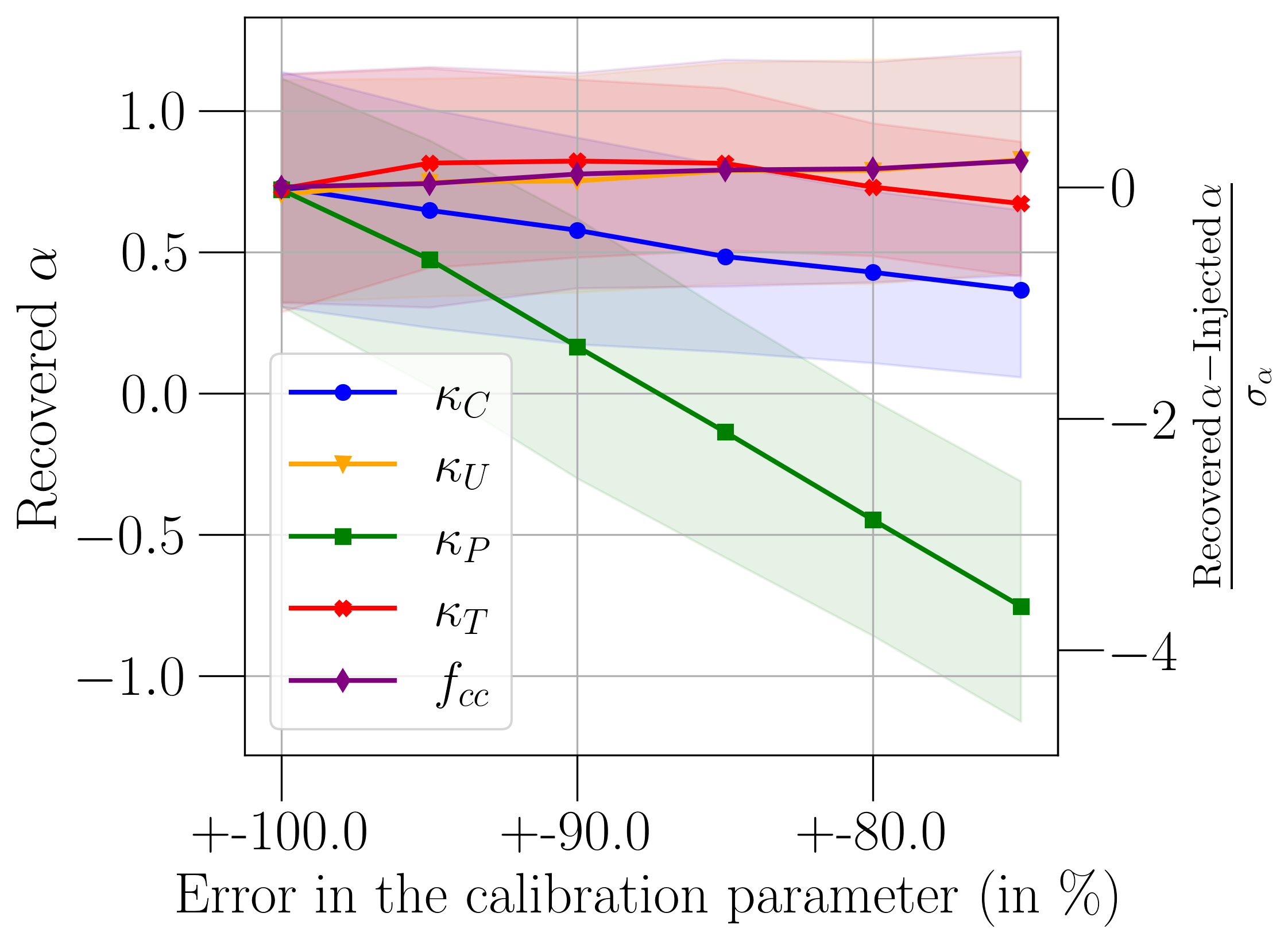} &
    \includegraphics[width=0.3\textwidth]{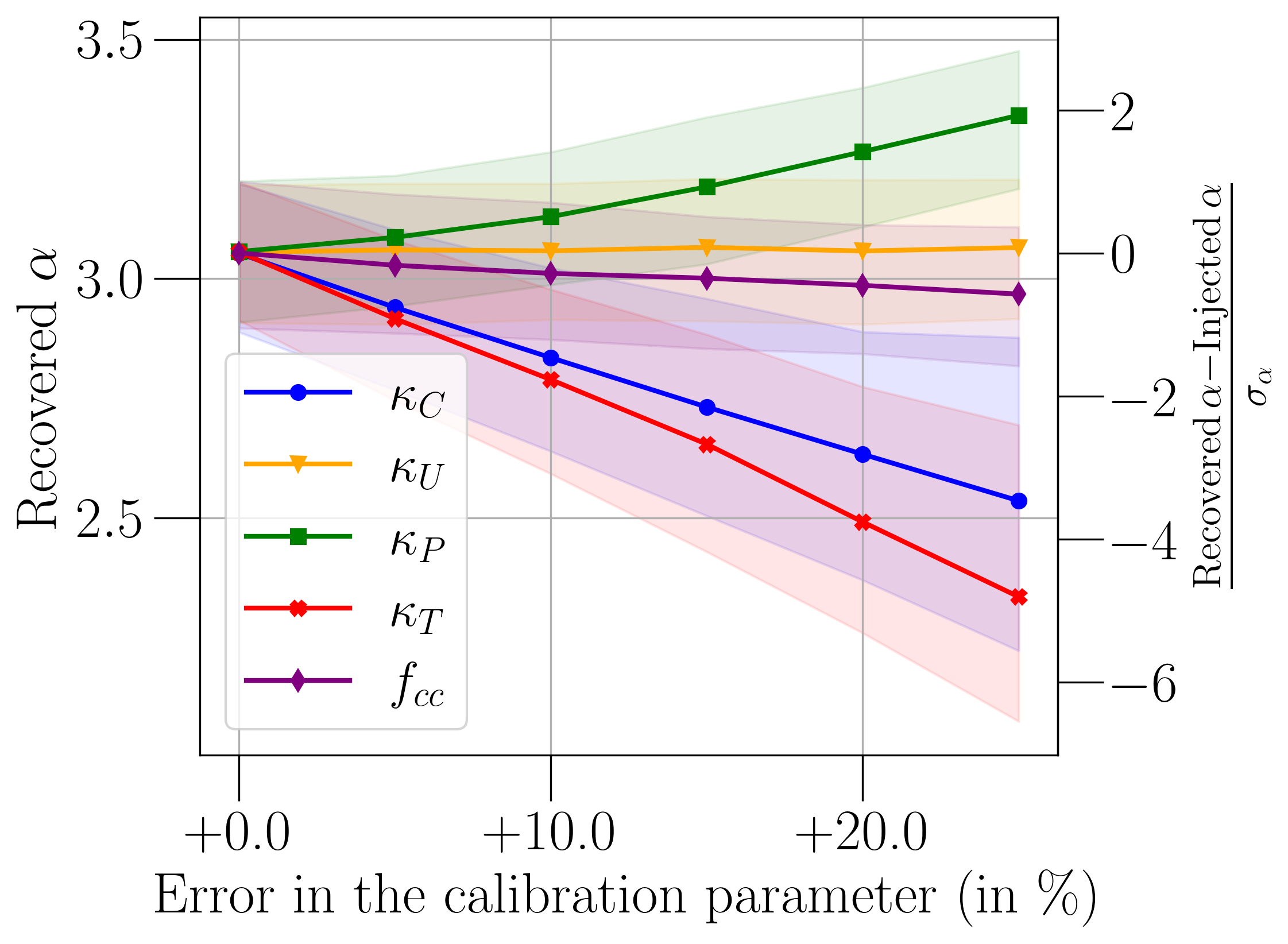}
  \end{tabular}
    \caption{Effect of the errors in various calibration model parameters on the recovery of the signal parameter $\Omega_{\alpha}$ and $\alpha$ for isotropic GWB signals described by $\alpha={0,2/3,3}$. The solid lines correspond to the maximum likelihood values, and the shaded regions indicate 68 \% confidence interval. The injected values of $\Omega_{\alpha}$ are $1.21 \times 10^{-8}$, $1.04 \times 10^{-8}$, and $2.70 \times 10^{-9}$  for $\alpha = 0 , 2/3, 3$ respectively.}
    \label{fig:maxL_values_errors}
\end{figure*}

The maximum likelihood values of the recovered parameters $\Omega_{\alpha}$ and $\alpha$, for $\alpha = 0, 2/3, 3$, as functions of errors on the various calibration parameters are shown in Fig.~\ref{fig:maxL_values_errors}. 
The plots in Fig.~\ref{fig:maxL_values_errors} show the recovered values of $\Omega_\alpha$ and $\alpha$ as we increase the errors on the calibration parameters $\kappa_C$, $f_{cc}$, $\kappa_U$, $\kappa_P$ and $\kappa_T$ in the response function $R(f)$ used to calibrate the detector output. For testing the recovery, we inject isotropic GWBs with amplitudes of $\Omega_\alpha = 1.21 \times 10^{-8}, 1.04 \times 10^{-8}, 2.70 \times 10^{-9}$ for $\alpha = 0 , 2/3, 3$ respectively and try to recover them with and without errors on the above calibration parameters. On the right side of the plots in Fig.~\ref{fig:maxL_values_errors}, we also show the difference between the injected and recovered values normalized by the 1-sigma uncertainties in the recovery. To have a common y-axis on the right side, for each $\alpha$, we use the largest 1-sigma uncertainty we observe among different calibration parameters for the normalization.

\begin{figure}
    \centering
    \includegraphics[width=\columnwidth]{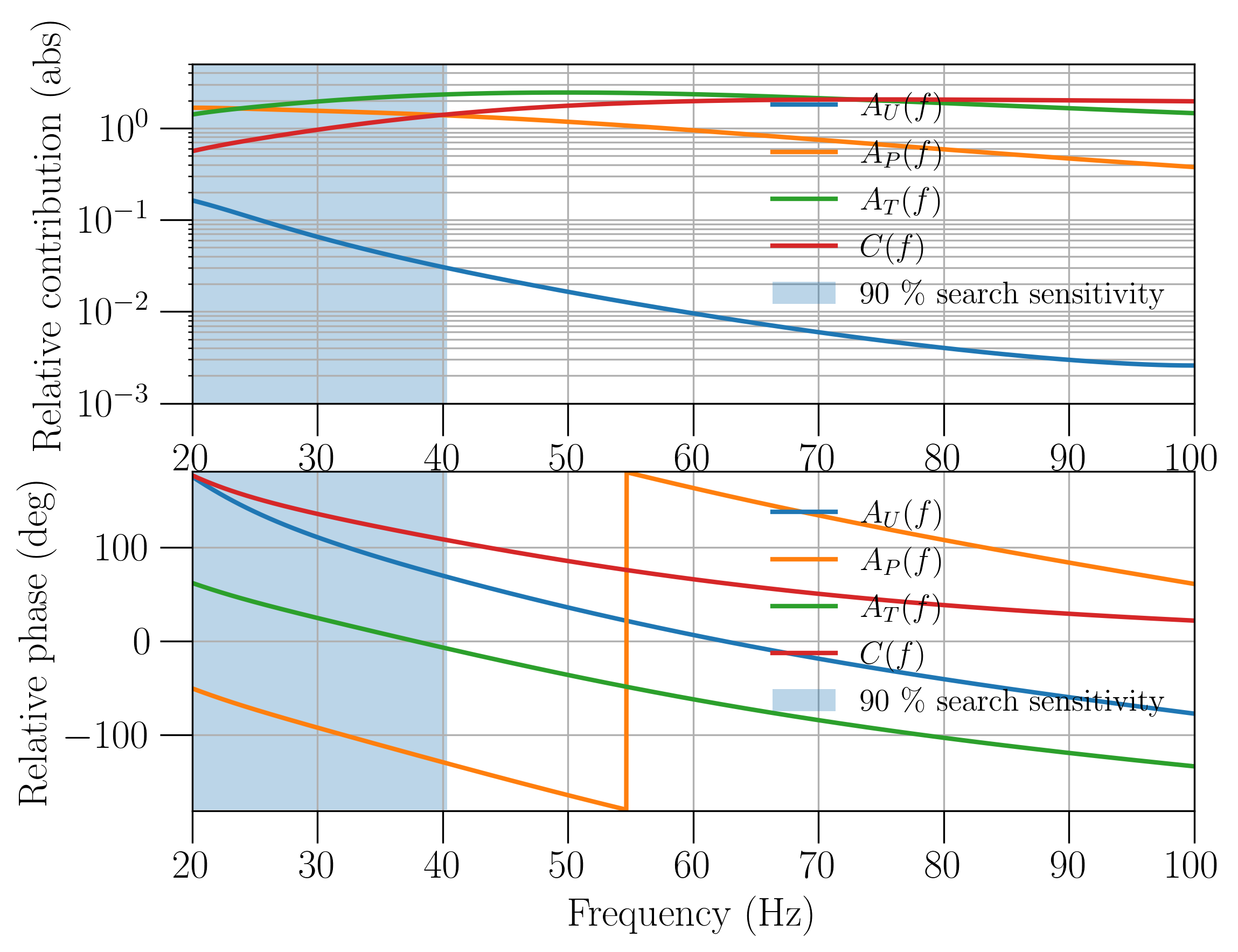}
    \caption{Relative contribution of various calibration parameters to the interferometer response function and 90 \% search sensitivity region for the $\alpha=2/3$ GWB search. For $\alpha=0$ and $\alpha=3$, the 90 \% search sensitivity regions extend up to $\approx 40$ Hz and $\approx 175$ Hz respectively. Because of the non-trivial phase relationship between different components in Eq.\ref{eq:relative_contribution}, we see that individual components' relative contributions to the response function can even go above one.}
    \label{fig:relative_contribution}
\end{figure}

We use the maximum likelihood method described in \cite{StochPE} and use {\tt dynesty} \cite{dynesty_paper} sampler in {\tt bilby} \cite{bilby_paper} package for sampling the likelihoods and estimating the maximum likelihood values of $\Omega_{\alpha}$ and $\alpha$ (shown in Fig.~\ref{fig:maxL_values_errors}) from $\hat{\Omega}_\alpha (f)$ and $\sigma_{\hat{\Omega}_\alpha}(f)$. From the plots in Fig.~\ref{fig:maxL_values_errors}, we see that when the errors on the calibration model parameters are zero, we recover the injected values very well. However, the recovered values of $\Omega_\alpha$ and $\alpha$ become biased as we increase the error on the calibration model parameters.  The errors on $\kappa_P$, $\kappa_T$ and $\kappa_C$ significantly bias the recoveries of $\Omega_{\alpha}$ and $\alpha$ while $f_{cc}$ and $\kappa_U$ have very little effect. For example, for $\alpha = 2/3$, with $10 \, \%$ error on the $\kappa_T$ the recovered $\Omega_\alpha$ is $\approx 2.5 \, \sigma_{\Omega_\alpha}$ away from its true value, while with $10 \, \%$ error on the $\kappa_P$ the recovered $\alpha$ is 
$\approx 1.5 \, \sigma_{\alpha}$ away from its true value. We also notice that, even though $\kappa_T$ significantly affects the $\Omega_\alpha$ estimate, it has minimal impact on the recovery of $\alpha$. These effects are likely due to how these different terms contribute to the interferometer response function. Rewriting Eq.~\ref{eq:response_function} into contributions from different components, we get,
\begin{eqnarray} \label{eq:relative_contribution}
   R(f) &=& 1/C(f) + \kappa_U D(f) A_U(f) \nonumber\\
        & & + \kappa_P D(f) A_P(f) + \kappa_T D(f) A_T(f) . 
\end{eqnarray}
Fig.~\ref{fig:relative_contribution} shows the relative contribution of the different terms in Eq.~\ref{eq:relative_contribution} to the response function and also 90 \% search sensitivity region for the $\alpha=2/3$ isotropic GWB. The 90 \% isotropic GWB search sensitivity region increases as we increase the values of $\alpha$. For $\alpha = 2/3$, the 90 \% search sensitivity region extends up to $\approx 45$ Hz, while for $\alpha=0$ and $\alpha=3$, the 90 \% search sensitivity regions extend up to $\approx 40$ Hz and $\approx 175$ Hz respectively. 

We see that in the 90 \% sensitivity region, penultimate and test mass actuation and sensing functions make the most significant contributions. The top test mass actuation function contributes $\lesssim 10 \%$ to the response function in the $20 - 1726$ Hz band and hence does not affect the signal recovery. In the sensing function (see Eq.\ref{eq:sensing_function}), the dominant contribution comes from $\kappa_C$. Since the typical value of $f_{cc}$ of advanced LIGO detectors during the O3 run was $\sim 400 \, {\rm Hz}$ and the 90 \% search sensitivity region extents only up to a maximum of $\sim 200$ Hz (for $\alpha = 3$), the effect of $f_{cc}$ on the estimation of the parameters is minimal. 

Since the $\alpha$ values of 0 and 2/3 are relatively closer, the results of $\alpha=0$ and $\alpha=2/3$ in Fig.~\ref{fig:maxL_values_errors} are very similar. We also observe that the result for $\alpha=3$ is slightly different. Since $\alpha =3$ probes a much larger frequency band of $\sim 20 - 175$ Hz where contributions from $\kappa_C$ and $\kappa_T$ to the response function tend to be larger on average compared to the other parameters (see Fig.~\ref{fig:relative_contribution}), $\kappa_C$ and $\kappa_T$ start to affect the recoveries of $\Omega$ and $\alpha$ significantly. We see this for $\alpha =3$ in Fig.~\ref{fig:maxL_values_errors}. 

The design of the detector, for example, the finesse of arm and recycling cavities, determines the cavity pole frequency, while the control architecture of the detector determines the relative contributions of different actuation stages. Thus, the effects of different calibration factors on the isotropic GWB search heavily depend on the detector's design and operation. 

\begin{figure*}[h]
  \centering
   \includegraphics[width=\textwidth]{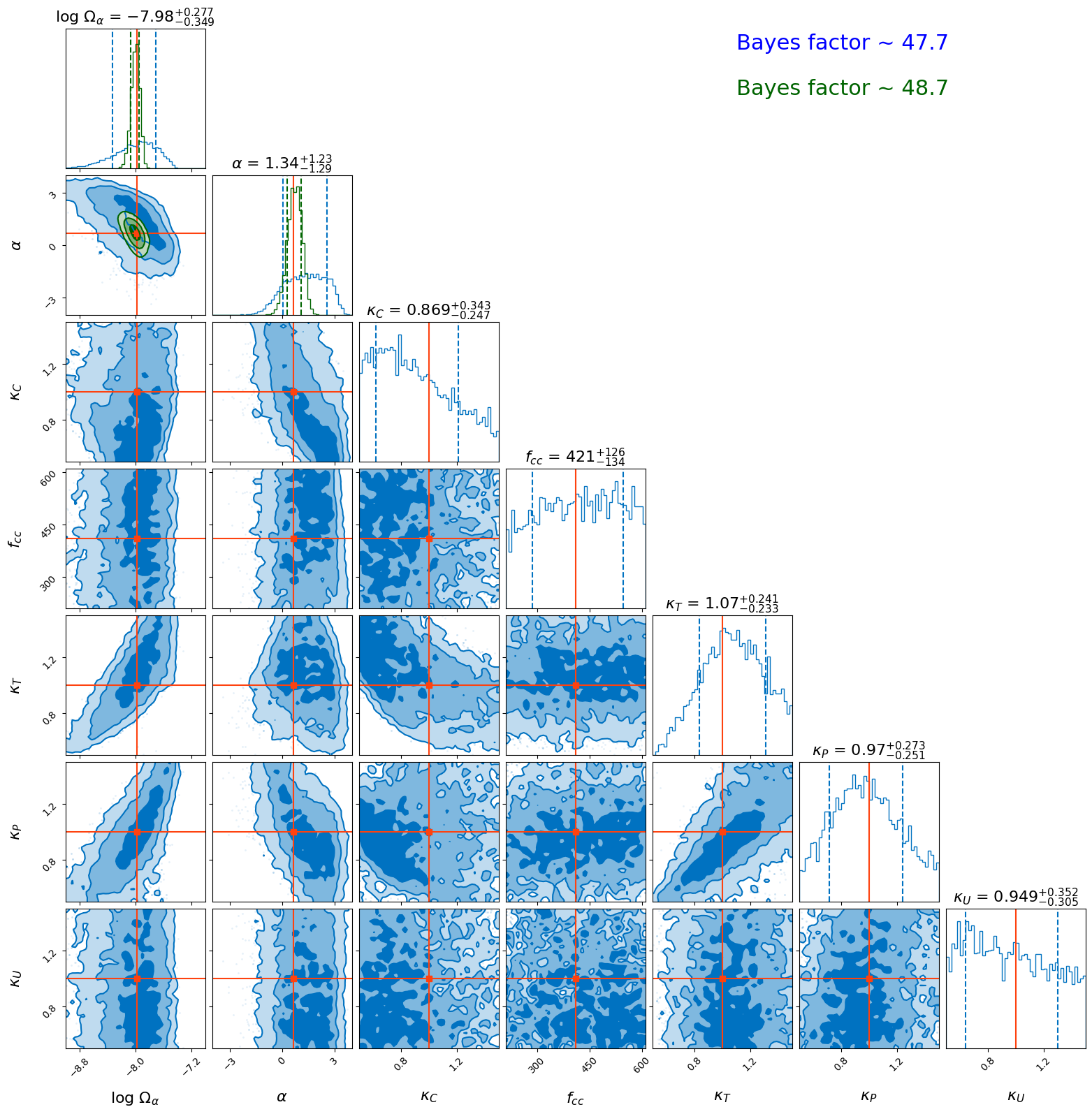}
    \caption{Corner plot (blue) showing the recovery of GWB and calibration model parameters using raw, uncalibrated data. For comparison, we also show the recovery of GWB parameters (green) using well-calibrated data. The injected values are shown using the (red) cross and vertical lines. The Bayes factors comparing signal vs. noise hypothesis for the two cases have also been shown.}
 \label{fig:corner_plot}
\end{figure*}

We also try to simultaneously estimate the calibration and GWB signal parameters to see how well we can do. Here we use (simulated) uncalibrated raw digital signals to extract all the parameters. Fig.~\ref{fig:corner_plot} shows an example of the simultaneous estimation of all the parameters for the $\alpha=2/3$ signal model. The plot shows that, along with the GWB model parameters, we can also infer the values $\kappa_P$, $\kappa_T$, and $\kappa_C$ to some level, but recoveries of $f_{cc}$ and $\kappa_U$ are poor which are consistent with the results in Fig.~\ref{fig:maxL_values_errors}. For comparison, we also show the recovery of GWB model parameters using calibrated data without any uncertainties. The plots also have the Bayes factors, comparing the signal vs. noise hypothesis for those two cases. We see that the Bayes factors do not change significantly in the two cases (as expected, it is slightly lower when we estimate calibration parameters also). However, the posteriors of GWB parameters are very broad and probably biased when we simultaneously estimate the GWB and calibration model parameters. So it is crucial to have well-calibrated data to get better posteriors on the signal parameters and a better Bayes factor. 

\section{Conclusions}\label{sec:conclusions}
 In this work, we have studied the effect of calibration uncertainties on the detection and parameter estimation of isotropic GWB signals. We focused on the amplitude ($\Omega_\alpha$) and power law index ($\alpha$) of power-law isotropic GWBs. We find that, for the second generation of gravitational wave detectors, when the calibration uncertainties are less than $\sim 10 \%$, they do not significantly affect the detection of a GWB signal. The calibration uncertainties of the LIGO detectors reported during the last observing run O3 are well within this $\sim 10 \%$ limit \cite{Sun_2020}. 

We also find that the recovery of isotropic GWB model parameters could be affected depending on which calibration parameter is poorly constrained and its uncertainty level.  The recovered values of signal parameters are biased due to errors in calibration model parameters. Even though the current errors on the individual model parameters of LIGO detectors are much smaller ($\lesssim 1 \%$), the cumulative effect of the different parameters could bias the recovered GWB parameters. Currently, this bias is not considered during the GWB parameter estimation or upper limit calculation. For a calibration uncertainty of $\sim 5$ \% of the interferometer response function (90 \% maximum reported for the LIGO detectors during O3), the biases in estimating GWB amplitudes or its upper limits are not significant $\lesssim 2$ \%. However, this might become significant for larger calibration uncertainties, especially when we try to differentiate between different models of GWB. In this work, we also try to estimate the isotropic GWB and calibration model parameters simultaneously and find that we could detect the GWB signal, albeit with some loss of Bayes factor (SNR). However, the posteriors of the GWB signal parameters become very broad and probably biased due to their correlation with some of the calibration parameters. This suggests the importance of well-calibrated data for detecting and recovering GWB signals, which is expected to be in the near future. 

We also note that the analysis presented in this paper highly depends on the GW detectors' calibration model (parameters). Hence, one might need to repeat this study when the calibration model changes significantly, for example, for future detectors. However, if the calibration uncertainties are kept small ($\lesssim 5 \%$), as we see in our analysis in this paper, the effects on the isotropic GWB analyses are expected to be small. Since the calibration model depends on the detector design and its control system architecture, one could also choose to design future detectors that would reduce the effect of calibration uncertainties. This is something that could be studied further. One could also extend the study reported in this paper to estimate the effect of calibration uncertainties on the GWB with more complicated model parameters or anisotropic GWB.

\section*{Acknowledgements}
The authors thank Jeffrey S Kissel for providing useful comments on the draft. The authors acknowledge the use of the IUCAA LDG cluster Sarathi for the 
computational/numerical work. J. Yousuf also acknowledges IUCAA for providing accommodation while carrying out this work. J. Yousuf is thankful to the Department of Science and Technology (DST), Government of India, for providing financial assistance through INSPIRE Fellowship. For this work, we used the software packages {\tt pyDARM} \cite{pydarm}, {\tt bilby} \cite{bilby_paper}, {\tt stochastic} \cite{stochastic_pipeline} and {\tt Matplotlib} \cite{matplotlib}.

\bibliographystyle{apsrev4-1}
\bibliography{GWBandCalibration}

\end{document}